\documentclass[aps,amsmath,amssymb,superscriptaddress,prb,10pt,twocolumn,eqrefnum,x11names]{revtex4-2}

\usepackage{graphicx}
\usepackage{dcolumn}
\usepackage{bm}
\usepackage{physics}
\usepackage{amsmath}
\usepackage{color}
\usepackage{hyperref}
\usepackage{float}
\usepackage{svg}
\usepackage{esvect}
\usepackage{comment}

\newcommand{\upup}{{\uparrow\uparrow}}
\newcommand{\dodo}{{\downarrow\downarrow}}

\newcommand{\rS}{{\mathrm{S}}}

\renewcommand{\vec}[1]{\textbf{#1}}

\newcommand{\red}[1]{\textcolor{red}{#1}}
\newcommand{\orange}[1]{\textcolor{orange}{#1}}
\newcommand{\lila}[1]{\textcolor{Orchid2}{#1}}
\newcommand{\green}[1]{\textcolor{Green3}{#1}}
\newcommand{\blue}[1]{\textcolor{blue}{#1}}

\begin{document}

\title{Purely even harmonic Josephson current due to crossed pair transmission across strongly spin-polarized materials}

\author{Niklas L. Schulz}
\email{niklas.schulz@uni-greifswald.de}
\affiliation{Institute of Physics, University of Greifswald, Felix-Hausdorff-Strasse 6, 17489 Greifswald, Germany}
\author{Danilo Nikoli\'c}
\email{danilo.nikolic@uni-greifswald.de}
\affiliation{Institute of Physics, University of Greifswald, Felix-Hausdorff-Strasse 6, 17489 Greifswald, Germany}
\author{Matthias Eschrig}
\email{matthias.eschrig@uni-greifswald.de}
\affiliation{Institute of Physics, University of Greifswald, Felix-Hausdorff-Strasse 6, 17489 Greifswald, Germany}

\date{\today}

\begin{abstract}
We revisit the problem of the second harmonic generation in the current-phase relation across ferromagnetic bilayers placed between BCS superconductors. In particular, we consider a strongly spin-polarized metallic ferromagnet coupled to two superconducting leads via thin spin-active (left) and non spin-active (right) insulating layers. The system is examined in the framework of the quasiclassical Green's function formalism both in the ballistic (Eilenberger) and the diffusive (Usadel) limit. Strong spin polarization allows for neglecting short-range mixed-spin correlations, and the Josephson supercurrent across the ferromagnet is fully mediated by long-range equal-spin triplet correlations. Using a diagrammatic technique for ballistic propagators introduced in Refs. \cite{zhaoDynamicsSpinTransport2007,zhaoTheoryNonequilibriumSpin2008,tomas2002}, we describe the relevant Andreev processes responsible for the effective conversion of two spin-singlet Cooper pairs in the superconductor into two $\upup$ and $\dodo$ pairs in the ferromagnet. Contrary to the naive picture of direct conversion, we show that the lowest order process involves four Cooper pairs in the superconductor, among which three are incoming and one is outgoing giving rise to net charge transport of $4e$ across the non spin-active interface. The self-consistent numerical treatment of the diffusive junction, typically more relevant in experiments, confirms this picture quantitatively.
\end{abstract}

\maketitle

\section{Introduction}
The understanding of long-range equal-spin triplet currents is of fundamental importance in superconducting spintronics~\cite{eschrigSingletTripletMixingSuperconductor2004,buzdinProximityEffectsSuperconductorferromagnet2005,bergeretOddTripletSuperconductivity2005,eschrigSpinpolarizedSupercurrentsSpintronics2011,eschrigSpinpolarizedSupercurrentsSpintronics2015,linderSuperconductingSpintronics2015,birgeSpintripletSupercurrentsJosephson2018,linderOddfrequencySuperconductivity2019,yangBoostingSpintronicsSuperconductivity2021,caiSuperconductorFerromagnetHeterostructures2023}. One of the most examined platforms for probing such currents are hybrids involving spin-singlet BCS superconductors (SC) and itinerant ferromagnets (FM), which can host pair amplitudes with contributions both from spin singlets and from spin triplets~\cite{tokuyasuProximityEffectFerromagnetic1988,demlerSuperconductingProximityEffects1997,bergeretLongRangeProximityEffects2001,eschrigSymmetriesPairingCorrelations2007,tanakaTheoryProximityEffect2007}. Using the exchange field of the ferromagnet as a global quantization axis allows for identifying mixed-spin correlations (spin singlets and spin triplets with $s_z=0$) and equal-spin triplet correlations (triplets with $s_z=\pm 1$), among which only the latter appear to be long-ranged~\cite{bergeretLongRangeProximityEffects2001}. This classification indicates that devices involving strongly spin-polarized ferromagnets (sFM) very effectively suppress the short-ranged mixed-spin correlations, such that long junctions almost exclusively host the long-ranged equal-spin triplet correlations and currents, leading to the emergence of new functionalities which make such setups particularly interesting~\cite{eschrigTheoryHalfMetalSuperconductor2003,eschrigTripletSupercurrentsClean2008,greinSpinDependentCooperPair2009,eschrigScatteringProblemNonequilibrium2009,greinInverseProximityEffect2013,eschrigSpinpolarizedSupercurrentsSpintronics2015,houzetQuasiclassicalTheoryDisordered2015,bobkovaGaugeTheoryLongrange2017,ouassouTripletCooperPairs2017,eschrigTheoryAndreevBound2018,schulzQuantumGeometricSpinCharge2025,schulzTheoryQuantumgeometricCharge2025,nikolicSpinresolvedJosephsonDiode2025}. The microscopic origin of long-range equal-spin correlations relies on two fundamental processes occurring close to the SC/sFM interface: (i) spin-mixing effect and (ii) spin-rotation mechanism~\cite{eschrigSpinpolarizedSupercurrentsSpintronics2011}, where the latter requires magnetic inhomogeneity in the system. The essential role of these two mechanisms in the creation and control of spin-triplet supercurrents in SC/FM hybrids has been confirmed  experimentally~\cite{keizerSpinTripletSupercurrent2006,anwarLongrangeSupercurrentsHalfmetallic2010,robinsonControlledInjectionSpinTriplet2010,khaireObservationSpinTripletSuperconductivity2010,lahabiControllingSupercurrentsTheir2017,glickPhaseControlSpintriplet2018,carusoTuningMagneticActivity2019,kapranObservationDominantSpintriplet2020,aguilarSpinpolarizedTripletSupercurrent2020}. In this paper, we show that once present, equal-spin triplets from sFM can be transmitted into a spin-singlet SC via higher-order Andreev processes, occurring at the interface, even without a spin-rotation mechanism. We classify the relevant Andreev processes and relate them to crossed pair transmission processes.


In general, Josephson junctions feature nonsinusoidal current-phase relations (CPRs) that can be decomposed as $I(\Delta\chi)=I_1\sin(\Delta\chi)+I_2\sin(2\Delta\chi)+\dots$, where $\Delta\chi$ is the superconducting phase difference and the $n$-th harmonic corresponds to the coherent transfer of $n$ Cooper pairs across the junction. The presence of a second harmonic in CPR plays an important role in various effects such as the Josephson diode effect~\cite{nadeemSuperconductingDiodeEffect2023,shafferTheoriesSuperconductingDiode2025} and stabilization of superconducting qubits~\cite{smithSuperconductingCircuitProtected2020,larsenParityProtectedSuperconductorSemiconductorQubit2020} as well as in superconducting digital electronics~\cite{mitrovicJosephsonJunctionsFerromagnetic2025a}. Considering junctions with ferromagnetic materials, it has been established that the second harmonic can be dominating the Josephson CPR. In junctions with uniform ferromagnets, this situation is realized at the $0-\pi$ transition due to the coexistence of the two phases~\cite{radovicCoexistenceStableMetastable2001}. However, this coexistence appears to be very sensitive to changes in the system's parameters~\cite{chtchelkatchevP0TransitionSuperconductorferromagnetsuperconductor2001,ryazanovCouplingTwoSuperconductors2001,barashInterplaySpindiscriminatedAndreev2002,sellierTemperatureinducedCrossover02003,radovicJosephsonEffectDoublebarrier2003,buzdinPeculiarPropertiesJosephson2005,houzetSuperharmonicJosephsonRelation2005}. In order to stabilize it, the presence of magnetic inhomogeneity that creates long-range equal-spin triplet correlations is necessary. 

The simplest realization of the Josephson junction with an inhomogeneous ferromagnet is a heterostructure involving a ferromagnetic bilayer with misaligned magnetizations~\cite{blanterSupercurrentLongSFFS2004,pajovicJosephsonCouplingFerromagnetic2006,crouzyJosephsonCurrentSuperconductorferromagnet2007,trifunovicLongRangeSuperharmonicJosephson2011,trifunovicJosephsonEffectSpintriplet2011,melnikovInterferencePhenomenaLongRange2012,richardSuperharmonicLongRangeTriplet2013,kawabataRobustnessSpinTripletPairing2013, hikinoLongRangeSpinCurrent2013, mengLongrangeSuperharmonicJosephson2016,mengJosephsonCurrentFerromagnetic2019,nikolicInterferencePhenomenaJosephson2022}.
{In Ref.~\cite{pajovicJosephsonCouplingFerromagnetic2006}, based on the solution of the Bogoliubov-de Gennes equation a strong second harmonic has been predicted to emerge across a ballistic ferromagnetic bilayer with orthogonal magnetizations. However, since a symmetric junction was considered its occurrence was attributed to the coexistence of the $0$ and $\pi$ phases, and its long-range nature was not clear. Based on the scattering approach, a junction comprising a metallic ferromagnet and a ferromagnetic insulating layer at the interface with a superconductor, was examined in Ref.~\cite{trifunovicLongRangeSuperharmonicJosephson2011}. It was explicitly shown that for long metallic ferromagnets, the odd harmonics are strongly suppressed as compared to the even ones. The long-ranged nature of the second harmonic was attributed to a transport process involving two normal and two anomalous Andreev reflections. Further quasiclassical studies of similar systems in the ballistic (Eilenberger) limit were reported in Refs.~\cite{trifunovicJosephsonEffectSpintriplet2011} and~\cite{melnikovInterferencePhenomenaLongRange2012}. The fully selfconsistent analysis in Ref.~\cite{trifunovicJosephsonEffectSpintriplet2011} showed that the long-range spin-triplet correlations are dominant in highly asymmetric junctions. In this case, the Josephson CPR hosts a strong second harmonic that is more robust against variation of the system's parameters in comparison to a second harmonic contribution induced by a $0$-$\pi$ transition. In Ref.~\cite{melnikovInterferencePhenomenaLongRange2012} an analytic study confirmed the findings of Ref.~\cite{trifunovicJosephsonEffectSpintriplet2011} emphasising the importance of the asymmetry of ferromagnetic bilayers for the second harmonic generation. In addition, it was shown that besides ferromagnetic bilayers, ferromagnetic wires with spin-orbit coupling can also host a long-range second harmonic in the CPR. A similar study in the diffusive (Usadel) limit was reported Ref.~\cite{richardSuperharmonicLongRangeTriplet2013}. As in the above works, it was shown that a high asymmetry of the ferromagnetic bilayer is demanded for the emergence of the long-range second harmonic. In addition, the authors proposed the idea that such contribution to the Josephson current relies on the coherent transport of two equal-spin triplet pairs with opposite spin orientation.}

{The mentioned studies showed the existence of a dominant second harmonic in the Josephson CPR across asymmetric ferromagnetic bilayers. However, a direct evidence that the current is mediated only by equal-spin triplet pairs was lacking. Furthermore, in Refs.~\cite{trifunovicJosephsonEffectSpintriplet2011,melnikovInterferencePhenomenaLongRange2012,richardSuperharmonicLongRangeTriplet2013} the quasiclassical condition restricts the exchange field strengths to rather small values, $J\lesssim 0.1 E_F$, which does not allow for complete neglect of mixed-spin correlations. Finally, a direct insight into the relevant Andreev processes was lacking, leaving the microscopic picture unclear.}

From an experimental point of view, an enhanced second harmonic in the long-range Josephson current-phase relation has been observed in mesa-heterostructures of cuprate superconductors and ferromagnetic bilayers of  manganite and ruthenate~\cite{ovsyannikovTripletSuperconductingCorrelations2013}, while the pure second harmonic has been observed in $\rm NbN/GdN/NbN$ junctions~\cite{palPureSecondHarmonic2014}. Note that odd harmonics in the CPR can be also long-ranged, but only in junctions with three or more ferromagnetic layers ~\cite{volkovOddTripletSuperconductivity2003,lofwanderInterplayMagneticSuperconducting2005,braudeFullyDevelopedTriplet2007,houzetLongRangeTriplet2007,fominovJosephsonEffectDue2007,sperstadJosephsonCurrentDiffusive2008,konschelleNonsinusoidalCurrentphaseRelation2008,greinSpinDependentCooperPair2009, volkovOddSpintripletSuperconductivity2010,alidoustSpintripletSupercurrentInhomogeneous2010,samokhvalovStimulationSingletSuperconductivity2014, haltermanChargeSpinCurrents2015,chenTripletProximityEffect2021,schulzQuantumGeometricSpinCharge2025,schulzTheoryQuantumgeometricCharge2025,nikolicSpinresolvedJosephsonDiode2025}. 

In this article, we show that a long Josephson junction involving a strongly spin-polarized ferromagnetic bilayer can host \textit{only} even-harmonic CPR. The strong spin polarization allows for a complete neglect of mixed-spin correlations, which in turn gives us an insight into the spin-resolved currents. Using a diagrammatic technique suited for ballistic quasiclassical propagators in the regime of strong spin polarization, we gain direct insight into the relevant Andreev processes. In particular, we obtain that the lowest order nonvanishing process contributing to the supercurrent involves two equal-spin triplet pairs in the ferromagnet and four singlet Cooper pairs in the superconductor yielding an effective transport of net charge of $4e$, consequently, giving rise to a dominant second harmonic in the CPR. This transmission process in which an effective transport of two singlet Cooper pairs as two equal-spin triplet pairs with the opposite spin orientations takes place is known as a \textit{crossed pair transmission} process \cite{greinSpinDependentCooperPair2009}. To confirm this picture, we numerically examine the same system in the diffusive limit, which is typically more relevant for experiments. 

\section{System and method} \label{sec:system}

\begin{figure}[htbp]
    \centering
    \includegraphics[width=0.9\linewidth]{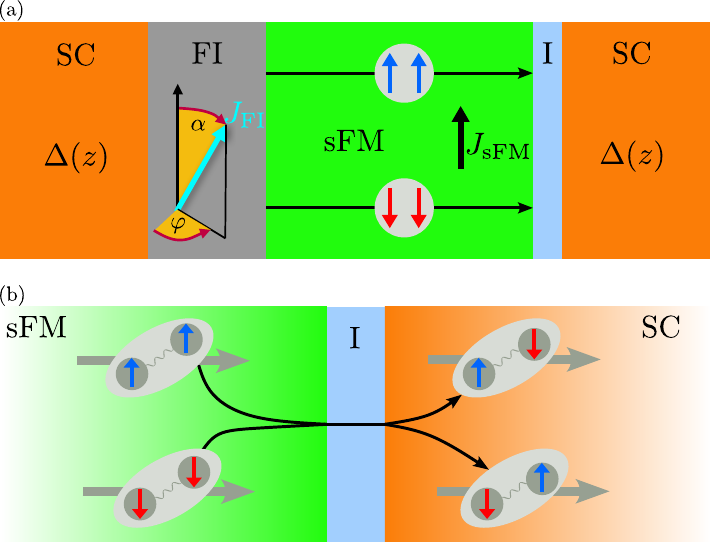}
    \caption{(a) Sketch of the system under study. A strongly spin-polarized metallic ferromagnet (sFM; green) is connected to two BCS superconductors (SC; orange) by a thin ferromagnetic insulator on the left side (FI; grey) and a thin non spin-active insulator on the right side (I; blue). The misalignment between the barrier's exchange field $\vec{J}_\mathrm{FI}$ and the exchange field in the sFM $\vec{J}_\mathrm{sFM}$, is characterized by spherical angles ($\alpha,\varphi$). (b) A schematic representation of an effective process that gives rise to the second harmonic in the CPR. Two equal-spin triplet pairs with opposite spin coming from the sFM are transmitted through the insulating barrier into the superconductor as two spin-singlet Cooper pairs. As we show in Sec.~\ref{sec:ballistic:results}, this is a naive representation showing just the effective process; the detailed Andreev scattering process is more complicated.}
    \label{fig:system_even_harmonics}
\end{figure}

We consider a system that consists of a strongly spin-polarized metallic ferromagnet (sFM) connected to two BCS superconductors (SC) by a thin ferromagnetic insulator (FI) on the left side and a thin normal insulator (I) on the right side [see Fig.~\ref{fig:system_even_harmonics}(a)]. The spin-rotational symmetry allows for choosing the exchange field of the sFM, $\vec{J}_\mathrm{sFM}$, as a global quantization axis. The exchange field of the left ferromagnetic insulator is assumed to point in an arbitrary direction w.r.t. $\vec{J}_\mathrm{sFM}$, $\vec{J}_\mathrm{FI}= J_\mathrm{FI} (\sin\alpha\cos\varphi,\sin\alpha\sin\varphi,\cos\alpha)$. Note that spin-rotational symmetry of the problem allows for setting $\varphi=0$, leaving $\alpha$ as the only relevant parameter in this respect. 

The metallic ferromagnet is strongly spin-polarized, i.e., $|\vec{J}_\mathrm{sFM}|\sim E_F$, where $E_F$ is the Fermi energy of the SC, yielding the suppression of mixed-spin correlations (spin singlets and spin triplets with $s_z=0$) on atomic length scales. As the size of the sFM exceeds these length scales by far, we neglect them inside the sample and account for equal-spin triplet correlations only (those with $s_z=\pm 1$). The insulating layers cannot be described within the quasiclassical theory, however, they enter the model via boundary conditions formulated in terms of the normal state scattering matrix~\cite{eschrigScatteringProblemNonequilibrium2009,eschrigGeneralBoundaryConditions2015}. In the following section, we outline the implementation of these assumptions, and we refer the reader for more details to Refs.~\cite{greinSpinDependentCooperPair2009,schulzQuantumGeometricSpinCharge2025,schulzTheoryQuantumgeometricCharge2025,nikolicSpinresolvedJosephsonDiode2025}. 

Next, we briefly discuss the relevant length and energy scales.
Superconductors are characterized by energy and length scales set by the superconducting transition temperature, $k_BT_c$, (or superconducting gap) and superconducting coherence length, which is given by $\xi_\mathrm{c} = \hbar v_F/k_BT_c$ in the clean, and $\xi_\mathrm{d} = \sqrt{\hbar D/k_BT_c}$ in the diffusive limit. Analogously, the metallic ferromagnet is characterized by an exchange field comparable to the Fermi energy, $E_F$, and ferromagnetic coherence length given by $\hbar v_{F,\eta}/k_BT_c$ in the clean, and by $\sqrt{\hbar D_{F,\eta}/k_BT_c}$ in the diffusive limit. In the expressions above, $v_F$ ($D$) refers to the Fermi velocity (diffusion constant) in the SC and $v_{F, \eta}$ ($D_\eta$) to the Fermi velocity (diffusion constant) of spin band $\eta = \uparrow,\downarrow$ in the sFM. Finally, the characteristic energy and length scales of the insulating layers are given by the Fermi energy $E_F$ and Fermi wave length, $\lambda_F$. In the results section we provide the specific values of the mentioned parameters, for which our simulations are performed. However, as we will show, the existence of the effect discussed in this paper is not sensitive to the exact values. 

For the microscopic description of the system, we make use of the quasiclassical theory of superconductivity, appropriate for treating mesoscopic systems both in and out of equilibrium~\cite{eilenbergerTransformationGorkovsEquation1968,larkinQuasiclassicalMethodTheory1969,belzigQuasiclassicalGreensFunction1999}. We restrict ourselves to an equilibrium situation and consider systems either in the ballistic or in the diffusive limit. These two regimes are realized for $\Delta_0\tau/\hbar\gg 1$ and $\Delta_0\tau/\hbar \ll 1$, respectively, where $\tau$ is the average electron-impurity scattering time and $\Delta_0$ is the superconducting gap at $T=0$. 

To gain insights in the underlying Andreev processes relevant for the Josephson transport, we first consider the ballistic junction described by the Eilenberger Green's function, $\hat{g}(\vec{p}_F,z,E)$~\cite{eilenbergerTransformationGorkovsEquation1968}, which allows for an analytic analysis. After that, we present the numerical results for the diffusive junction, which is typically more relevant for experiments, and which is described by the Usadel Green's function, $\hat{G}(z,E)$~\cite{usadelGeneralizedDiffusionEquation1970}. As we discuss below, in this case the orbital degree of freedom is integrated out, i.e., $\hat{G}(z,E)=\expval{\hat{g}(\vec{p}_F,z,E)}_{\vec{p}_F}$, where the averaging over the Fermi surface is assumed. 

\section{Ballistic limit} \label{sec:ballistic}
We first consider the junction in the ballistic limit as it allows for an analytic study of the relevant Andreev processes that contribute to the Josephson current. The study relies on the diagrammatic technique developed in Refs.~\cite{zhaoDynamicsSpinTransport2007,zhaoTheoryNonequilibriumSpin2008,tomas2002,eschrigScatteringProblemNonequilibrium2009} and elaborated in more detail in Appendix~\ref{sec:app:diagrams}. In what follows, we first present the microscopic description of each constituent of the junction, then we present the results.

\subsection{Theory} \label{sec:theory:ballistic}
\subsubsection{Superconductor} \label{sec:theory:ballistic:SC}
Mesoscopic superconductors in the ballistic limit can be described by the Eilenberger equation in the following form \cite{eilenbergerTransformationGorkovsEquation1968,larkinQuasiclassicalMethodTheory1969}:
\begin{equation}\label{eq:eilenberger}
    i\hbar \vec{v}_F\cdot\grad_{\vec{R}} \hat{g} + \comm{E \hat{\tau}_3 - \hat{\Delta}}{\hat{g}}  = \hat{0},
\end{equation}
supplemented by the normalization condition $\hat{g}^2 = \hat{1}$. Here, $E$ is the quasiparticle energy, $\hat{\tau}_3 = \mathrm{diag}(\mathit{1},-\mathit{1})$ is the third Pauli matrix with $\mathit{1}$ being the identity matrix in spin space,  $\hat{\Delta}$ is the BCS self-energy, which for a singlet superconductor has the form $\hat{\Delta} = \text{antidiag}(\Delta i\sigma_2,\Delta^*i\sigma_2)$, and $\vec{v}_F$ is the Fermi velocity. The Green's function itself has the following matrix structure in particle-hole $\otimes$ spin space:
\begin{equation}
    \hat{g} = \begin{pmatrix} g & f \\ -\tilde{f} & -\tilde{g} \end{pmatrix},\quad\text{with}\quad g = \begin{pmatrix} g_{\upup} &  g_{\uparrow\downarrow} \\ g_{\downarrow\uparrow} & g_{\dodo} \end{pmatrix},
\end{equation}
where the particle-hole conjugation operation $(\tilde{\,})$ is given by $\tilde{Q}(\vec{p}_F,\vec{R},E) = Q^\ast(-\vec{p}_F,\vec{R},-E^\ast)$. The normalization condition mentioned above allows us to parameterize the GF by the so-called Riccati coherence amplitudes $\gamma$ and $\tilde{\gamma}$ as follows \cite{eschrigDistributionFunctionsNonequilibrium2000,eschrigScatteringProblemNonequilibrium2009}:
\begin{align}
    g &= \qty(\mathit{1} - \gamma\tilde{\gamma})^{-1} \qty(\mathit{1} + \gamma\tilde{\gamma}), \label{eq:G}\\
    f &= \qty(\mathit{1}-\gamma\tilde{\gamma})^{-1} 2\gamma. \label{eq:F}
\end{align}
{The coherence amplitudes represent conversion processes between holelike and particlelike Andreev amplitudes, where $\gamma$ describes the conversion of holelike to particlelike Andreev amplitudes and $\tilde{\gamma}$ describes the conversion of particlelike to holelike Andreev amplitudes. In the case of a superconductor they are 2$\times$2 matrices.} Having obtained the Green's function, the charge current can be calculated as 
\begin{equation}
    \bm{j}(\vec{R}) = -eN_F \Re \! \int_{-\infty}^\infty \frac{dE}{4} \bm{j}_s(\vec{R},E)\tanh(\frac{E}{2k_BT}), \label{eq:charge_current_SC}
\end{equation}
where the spectral current $\bm{j}_s$ is given by
\begin{equation}\label{eq:j_S}
    \bm{j}_s(\vec{R},E)=\Tr_4 \langle \vec{v}_F \hat{\tau}_3 \hat{g}(\vec{p}_F,\vec{R},E) \rangle_{\vec{p}_F}.
\end{equation}
In the expressions above, $e = -\abs{e}$ is the electron charge, $\Tr_4$ denotes the trace over the combined particle-hole $\otimes$ spin space, $N_F$ is the density of states at the Fermi level, and $\langle \ldots \rangle_{\vec{p}_F}$ denotes averaging over the Fermi surface (FS)
\begin{align}
    \expval{\bullet}_{\vec{p}_F} &= \frac{1}{N_{F}}\int_\mathrm{FS}\frac{d^2p_{F}}{(2\pi\hbar)^3|\vec{v}_{F}(\vec{p}_{F})|}(\bullet),\label{eq:avegraging}\\
    N_{F} &= \int_\mathrm{FS}\frac{d^2p_{F}}{(2\pi\hbar)^3|\vec{v}_{F}(\vec{p}_{F})|}.\label{eq:DOS}
\end{align}
Rewritten in terms of Riccati amplitudes, the trace in Eq.~\eqref{eq:j_S} reads
\begin{equation}\label{eq:fs_to_gammas}
\begin{split}
    \Tr_4\{\hat{\tau}_3 \hat{g}\} &= 2\Tr_2\qty[\qty(\mathit{1}-\gamma\tilde{\gamma})^{-1} + \qty(\mathit{1}-\tilde{\gamma}\gamma)^{-1} - \mathit{1}]=\\
    &= 2\Tr_2 \qty[ \mathit{1} + \sum_{k=1}^\infty (\gamma\tilde{\gamma})^k + (\tilde{\gamma}\gamma)^k], 
\end{split}
\end{equation}
where $\Tr_4$ denotes the trace over the combined particle-hole $\otimes$ spin space and $\Tr_2$ is the trace only over spin space. These formulas lay the basis for the diagrammatic description of how the Andreev processes occurring at the SC/sFM interface contribute to the Josephson current. 

\subsubsection{Strongly spin-polarized ferromagnet} \label{sec:ballistic:sFM}
As anticipated in Sec.~\ref{sec:system}, we consider composite systems involving strongly spin-polarized ferromagnetic materials which neglect mixed-spin correlations. Due to strong splitting between the spin bands, the quantum coherence is maintained only within each band, but not between the two. Consequently, the correlations that involve both spin bands are strongly suppressed. This situation allows for performing the quasiclassical approximation for each of the two spin bands separately (for more details, see Refs.~\cite{greinSpinDependentCooperPair2009,eschrigScatteringProblemNonequilibrium2009,greinInverseProximityEffect2013,bobkovaGaugeTheoryLongrange2017,nikolicSpinresolvedJosephsonDiode2025}). Under these assumptions, the sFM is described by two quasiclassical propagators $\breve{g}_{\eta\eta}$, one for each spin projection $\eta$, obeying the following scalar Eilenberger equation and normalization condition:
\begin{equation}
     i\hbar \vec{v}_{F,\eta} \cdot \grad_{\vec{R}} \breve{g}_{\eta\eta} + \comm{E\breve{\tau}_3}{\breve{g}_{\eta\eta}}  = \breve{0}, \quad \breve{g}_{\eta\eta}^2 = \breve{1}. \label{eq:Eilenberger_sFM}
\end{equation}
The Green's function has the following matrix structure:
\begin{equation}\label{eqn:G_sigma}
    \breve{g}_{\eta\eta} = \begin{pmatrix} g_{\eta\eta} & f_{\eta\eta} \\ -\tilde{f}_{\eta\eta} & -\tilde{g}_{\eta\eta} \end{pmatrix},
\end{equation}
where, in contrast to the GF in the superconducting region, each entry is spin-scalar. 

Analogously to Eqs.~\eqref{eq:G} and~\eqref{eq:F}, the above propagator can be parametrized as follows: 
\begin{align}
    g_{\eta\eta} &= \qty(1 - \gamma_{\eta\eta}\tilde{\gamma}_{\eta\eta})^{-1}\qty(1 + \gamma_{\eta\eta}\tilde{\gamma}_{\eta\eta}), \label{eq:G_F}\\
    f_{\eta\eta} &= \qty(1-\gamma_{\eta\eta}\tilde{\gamma}_{\eta\eta})^{-1} 2\gamma_{\eta\eta}. \label{eq:F_F} 
\end{align}
The introduced spin-scalar coherence amplitudes obey the following Eilenberger equation [see Eq.~\eqref{eq:Eilenberger_sFM}]:
\begin{equation}
i\hbar\vec{v}_{F,\eta}\cdot\bm{\nabla}\gamma_{\eta\eta} + 2E\gamma_{\eta\eta}=0,  
\end{equation}
which is readily solved
\begin{equation}\label{eqn:gamma_F}
    \gamma_{\eta\eta}(E,z)=A_{\eta\eta} e^{2iEz/\hbar v_{F,\eta}}.
\end{equation}
Here, $A_{\eta\eta}$ is an integration constant to be determined from the boundary condition. Note that we assume $\vec{v}_{F,\eta}=v_{F,\eta}\vec{e}_z$.

The charge supercurrent is calculated from Eq.~\eqref{eq:charge_current_SC}, which now takes the form
\begin{equation}
    \bm{j}(\vec{R})\!=\!-2e\!\sum_{\eta=\uparrow,\downarrow}\int\limits_{-\infty}^\infty dE \bm{j}_{\eta\eta}(\vec{R},E)\tanh(\frac{E}{2k_BT}), \label{eq:ballistic_current}
\end{equation}
where the spectral current $\bm{j}_{\eta\eta}$ in terms of coherence amplitudes reads
\begin{equation}
   \bm{j}_{\eta\eta}(E,\vec{R})\!=\!\frac{N_{F,\eta}}{4}\!\expval{\! \vec{v}_{F,\eta} \frac{1+\gamma_{\eta\eta}\tilde{\gamma}_{\eta\eta}}{1-\gamma_{\eta\eta}\tilde{\gamma}_{\eta\eta}}}_{\vec{p}_{F,\eta}}.
\end{equation}
Note that $\gamma_{\eta\eta} = \gamma_{\eta\eta}(\vec{p}_{F,\eta},E,\vec{R})$, $N_{F,\eta}$ is the spin-resolved density of states at the corresponding Fermi level, and the averaging is performed over the Fermi surface with a full analogy to Eqs.~\eqref{eq:avegraging} and~\eqref{eq:DOS}.

\subsubsection{Boundary conditions} \label{sec:ballistic:bc}
As already discussed, we are interested in the microscopic origin of even harmonics in the current-phase relation across the sFM layer. According to the general theory, such contributions are mediated by even number of Cooper pairs. Since the left insulating layer is ferromagnetic the misalignment of its exchange field ($\vec{J}_\mathrm{FI}$) and that of the metallic ferromagnet ($\vec{J}_\mathrm{sFM}$) allows for the transmission of an arbitrary number of equal-spin pairs due to the spin-rotation mechanism. Thus, the restriction to the transport of only even number of Cooper pairs must originate from the right, purely insulating, layer between the sFM and the SC. Consequently, the boundary conditions imposed at this interface play a crucial role in this respect. For this purpose, we follow Ref.~\cite{eschrigScatteringProblemNonequilibrium2009} where the boundary conditions are formulated in terms of the microscopic normal-state scattering matrix (S-matrix) and are applied to the Riccati amplitudes introduced earlier. The core of this approach is to distinguish between the incoming (Fermi velocity pointing towards the interface) and outgoing (Fermi velocity pointing away from the interface) amplitudes which are coupled by the the S-matrix at the interface. Following the notation of Ref.~\cite{eschrigScatteringProblemNonequilibrium2009}, we denote the former by small letters, $\gamma$ and $\tilde{\gamma}$, and the latter by the capital ones, $\Gamma$ and $\tilde{\Gamma}$ (see also Refs.~\cite{eschrigDistributionFunctionsNonequilibrium2000,zhaoTheoryNonequilibriumSpin2008}). Keeping this in mind, we introduce renormalized Riccati amplitudes, which incorporate elementary scattering processes occurring at the interface, defined in channel space as
\begin{equation}
    \gamma_{ij}^\prime = \sum_{k=1,2,3} \bm{S}_{ik} \gamma_{k} \tilde{\bm{S}}_{kj}, \label{eq:elementary_scatter_process}
\end{equation}
where $\bm{S}_{ij}$ is the scattering matrix element allowing for transmission/reflection from $j$ to $i$. We use the convention that $i=1$ denotes the doubly spin degenerated band in $\mathrm{SC}$, and $i=2,3$ denotes the $\uparrow\!/\!\downarrow$-spin band in sFM. Since the right insulating layer is normal there are no spin-flip reflections from one spin-band to the other, i.e., $\bm{S}_{23} = \bm{S}_{32} = 0$. The outgoing coherence amplitude on the superconducting side $\Gamma_1$ follows as \cite{eschrigScatteringProblemNonequilibrium2009}
\begin{equation}
\Gamma_1 = \gamma_{11}^\prime + \Gamma_{1\leftarrow 2}\, \tilde{\gamma}_2\, \gamma_{21}^\prime + \Gamma_{1\leftarrow 3}\, \tilde{\gamma}_3\, \gamma_{31}^\prime. \label{eq:outgoing_1}
\end{equation}
\begin{figure*}[t]
    \centering
    \includegraphics[width=\linewidth]{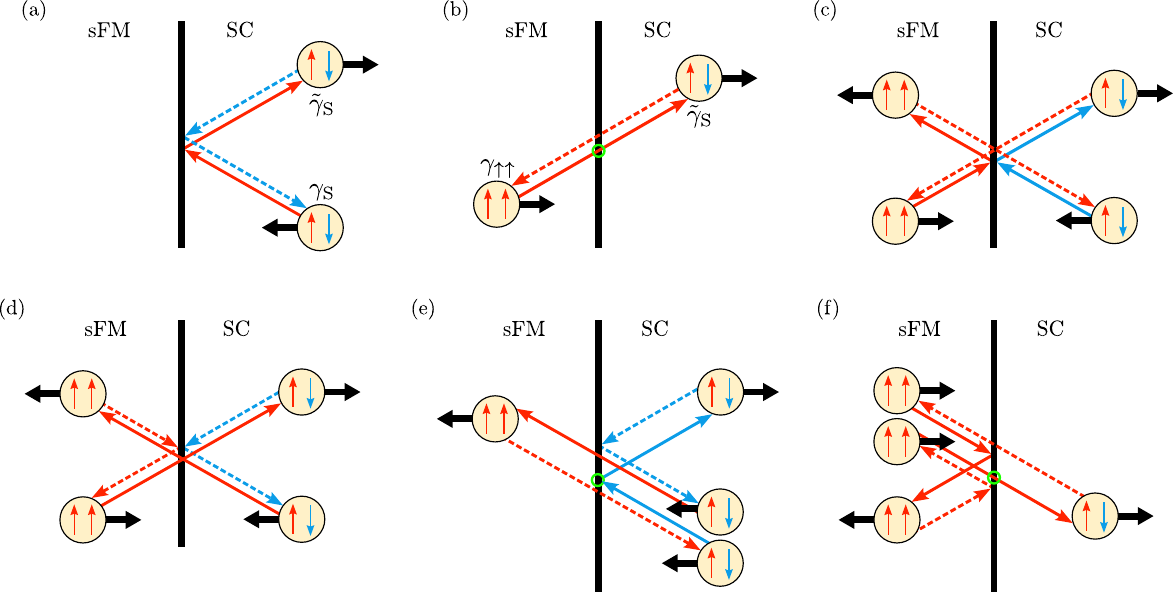}
    \caption{Sketches of the Andreev processes up to the second order in the transmission amplitude following from Eq.~\eqref{eq:Gamma_1} and explicitly shown in Eqs.~\eqref{eq:contribution_first} and \eqref{eq:contribution_third}. {The singlet Cooper pairs (coherence amplitudes $\gamma_{\rS},\tilde{\gamma}_{\rS}$) in the SC and the triplet pair correlations (coherence amplitudes $\gamma_{\eta\eta},\tilde{\gamma}_{\eta\eta}$) in the sFM are denoted by circles containing the corresponding spins. Red and blue colors denote spin-$\uparrow$ and spin-$\downarrow$, respectively. Here $\gamma_i$ describes the conversion process of holelike to particlelike Andreev amplitudes whereas $\tilde{\gamma}_i$ describes the conversion of particlelike to holelike Andreev amplitudes.} The black line in the middle of all panels represents the sFM/SC barrier which does not allow for spin-flip processes. The colored arrows represent the trajectory of a particle (solid line) or hole (dashed line). The black arrows at the edge of the singlet Cooper pairs or triplet pairs show the direction of the center-of-mass momentum. The green dot in panels (b), (e) and (f) indicate where a spin-flip process would be needed in order for the diagram to contribute to the charge current. All of the above diagrams also appear with spins exchanged, i.e., the red and blue color are exchanged.}
    \label{fig:lower_order_contributions}
\end{figure*}
\subsection{Results} \label{sec:ballistic:results}
In this section, we consider Andreev processes occurring at the right SC/sFM interface, which, as we show below, lead to a purely even-harmonic CPR. As mentioned in the previous section, the momentum dependence of the GF requires that we must distinguish between incoming and outgoing coherence amplitudes. Consequently, for Fermi velocities pointing away from the interface the trace in the expression for the supercurrent [see Eq.~\eqref{eq:charge_current_SC}] is rewritten as
    \begin{equation}
    \Tr_4\{\hat{\tau}_3 \hat{g}\} = 2\Tr_2\qty[ \mathit{1} + \sum_{k=1}^\infty (\Gamma\tilde{\gamma})^k + (\tilde{\gamma}\Gamma)^k]. \label{eq:trace_outgoing_gammas}
\end{equation}
The contributions for Fermi velocities pointing towards the interface are obtained by the replacement $\Gamma \rightarrow \gamma$ and $\tilde{\gamma} \rightarrow \tilde{\Gamma}$. Thus, we must consider the term $\Gamma \tilde{\gamma}$ to gain insights into the relevant processes. An intuitive way of visualising these microscopic processes is given by a diagrammatic notation introduced in Refs.~\cite{zhaoDynamicsSpinTransport2007,zhaoTheoryNonequilibriumSpin2008,tomas2002,eschrigScatteringProblemNonequilibrium2009}.

To build a diagrammatic theory, we first express the outgoing coherence amplitude [see Eq.~\eqref{eq:outgoing_1}] in terms of the incoming ones, $\gamma_i$ and $\tilde{\gamma}_i$, and the renormalized amplitudes, $\gamma_{ij}^\prime$, (for details, see Appendix~\ref{app:general_solution}):
\begin{align} \label{eq:Gamma_1}
        \Gamma_1 &= \gamma_{11}^\prime + \gamma_{12}^\prime \tilde{\gamma}_2 \gamma_{21}^\prime + \gamma_{13}^\prime \tilde{\gamma}_3 \gamma_{31}^\prime + \gamma_{13}^\prime \tilde{\gamma}_3 \gamma_{32}^\prime \tilde{\gamma}_2 \gamma_{21}^\prime \\
            &\quad + \gamma_{12}^\prime \tilde{\gamma}_2 \gamma_{23}^\prime \tilde{\gamma}_3 \gamma_{31}^\prime + \gamma_{13}^\prime \tilde{\gamma}_3 \gamma_{32}^\prime \tilde{\gamma}_2 \gamma_{23}^\prime \tilde{\gamma}_3 \gamma_{31}^\prime + \Gamma^\mathrm{higher}_1.\nonumber
\end{align}
Here, $\Gamma^\mathrm{higher}_1$ contains higher order contributions which are explicitly discussed in Appendix~\ref{app:general_solution}, where we show that these higher order contributions can be directly related to the ones explicitly shown in the above equation. Finally, we express the renormalized coherence amplitudes in terms of $\gamma_i$'s by using Eq.~\eqref{eq:elementary_scatter_process}. We do not show these expressions here but in Appendix~\ref{sec:app:diagrams}, see Eqs.~\eqref{eq:contribution_first}-\eqref{eq:contribution_seventh}. These expressions can then be directly interpreted in terms of reflection, transmission and Andreev reflection processes, and in the following we focus on the diagrammatic representation of $\Gamma_1 \tilde{\gamma}_1$. Before we comment on it, we outline the diagrammatic rules used for solving the scattering problem. 

The diagrams discussed in the following consist of solid/dashed and red/blue arrows. These arrows denote a trajectory, where the line style refers to a particle (solid line) or a hole (dashed line). The spin of the excitation is encoded in the color, red for spin-$\uparrow$ and blue for spin-$\downarrow$. The black line in the middle of each diagram represents the normal insulating layer, which allows for the particle (hole) to either undergo normal reflection or transmission but, importantly, with no spin-flip. The beige circles containing a red $\uparrow$ and a blue $\downarrow$ on the superconducting side represent singlet pair correlations (coherence amplitudes $\gamma_\rS,\tilde{\gamma}_\rS$). The black arrows refer to the direction of motion of the created Cooper pair, which is towards or away from the interface depending whether the hole or the particle is Andreev reflected. A similar notation is used for representing triplet pair correlations (coherence amplitudes $\gamma_{\eta\eta},\tilde{\gamma}_{\eta\eta}$) ferromagnetic side of the barrier. Note that we use the notation within a missing electron in spin-band $\eta$ is equivalent to a hole in spin-band $\bar{\eta}$, i.e., the Andreev reflected particle/hole on the superconducting side has the opposite spin to the incoming hole/particle. On the contrary, in the ferromagnet only equal-spin Andreev reflection is possible and thus the outgoing particle/hole has the same spin as the incoming hole/particle. {The coherence amplitude $\gamma_i$ ($\tilde{\gamma}_i$) describes the conversion process of holelike (particlelike) to particlelike (holelike) Andreev amplitudes.} The meaning of green circles at the barrier in some diagrams is explained below. 

\begin{figure}[t!]
    \centering
    \includegraphics[width=0.8\linewidth]{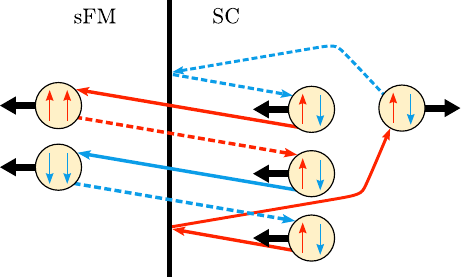}
    \caption{A schematic sketch of the lowest order processes contributing to the Josephson current across the junction following from Eq.~\eqref{eq:current_contribution}. The notation is the same as for Fig.~\ref{fig:lower_order_contributions}. This process leads to a net charge transfer of $4e$ and consequently a second harmonic contribution in the Josephson CPR.}
    \label{fig:lowest_current_contribution}
\end{figure}

{The technique outlined above is used to discuss the diagrammatic representation of the contributions to $\Gamma_1\tilde{\gamma}_1$. $\Gamma_1$ represents the coherence amplitude for a reteroreflection at the right SC/sFM interface of a holelike quasiparticle (coming from the SC) to a particlelike one. On the other hand, $\tilde{\gamma}_1$ is the coherence amplitude for a reteroreflection at the condensate leading to the conversion of a particlelike excitation (moving away from the interface) to a holelike excitation. The combination of $\Gamma_1 \tilde{\gamma}_1$ results in closed diagrams [see Figs.~\ref{fig:lower_order_contributions}, \ref{fig:lowest_current_contribution}, \ref{fig:fourth_order_Andreev_processes}, and \ref{fig:sixth_order_Andreev_processes}]. In contrast, the diagrams shown in Figs.~\ref{fig:app:normalisation_diagrams} and \ref{fig:app:normalisation_second_order}, discussed in Appendix~\ref{app:general_solution}, are not closed as they describe insertions for closed diagrams. Following these rules, the lowest order contributions to $\Gamma_1 \tilde{\gamma}_1$ [see Eqs.~\eqref{eq:contribution_first} - \eqref{eq:contribution_third}] are diagrammatically shown in Fig.~\ref{fig:lower_order_contributions}.}

The necessary condition for a process to contribute to the supercurrent is that a net charge is transferred across the interface. Thus, the number of incoming and outgoing pairs on the two sides of the interface must differ. This implies that the diagrams shown in Figs.~\ref{fig:lower_order_contributions}(a),~\ref{fig:lower_order_contributions}(c), and 
~\ref{fig:lower_order_contributions}(d) cannot contribute to the current as there is no net transport of electrons from one side of the barrier to the other. The remaining diagrams in Fig.~\ref{fig:lower_order_contributions}, i.e., panels (b), (e) and (f), do not contribute either. The reason for this is that the initial incoming hole (from the right moving Cooper pair in the SC) and the outgoing particle have the same spin. As we consider $s$-wave superconductors Andreev reflection only allows for a conversion between particles and holes with opposite spin (in the notation used here). Thus, an additional, here absent, spin-flip process is required for those processes to contribute to the supercurrent. A demand of such process for diagrams to contribute is denoted by a green circle in panels in Figs.~\ref{fig:lower_order_contributions}(b),~\ref{fig:lower_order_contributions}(e), and~\ref{fig:lower_order_contributions}(f). Therefore, all contributions up to the second order in the transmission amplitude do not contribute to the supercurrent, and we need to seek for higher orders in the expansion~\eqref{eq:Gamma_1}. Note that if spin-flip processes were present, the effective transport of one Cooper pair would be possible giving rise to a first harmonic in the CPR, which is in accordance with Refs.~\cite{volkovOddTripletSuperconductivity2003,lofwanderInterplayMagneticSuperconducting2005,braudeFullyDevelopedTriplet2007,houzetLongRangeTriplet2007,fominovJosephsonEffectDue2007,sperstadJosephsonCurrentDiffusive2008,konschelleNonsinusoidalCurrentphaseRelation2008,greinSpinDependentCooperPair2009, volkovOddSpintripletSuperconductivity2010,alidoustSpintripletSupercurrentInhomogeneous2010,samokhvalovStimulationSingletSuperconductivity2014, haltermanChargeSpinCurrents2015,chenTripletProximityEffect2021,schulzQuantumGeometricSpinCharge2025,schulzTheoryQuantumgeometricCharge2025,nikolicSpinresolvedJosephsonDiode2025}. 

{The first relevant contributions to $\Gamma_1$ follow from}
\begin{equation}
    \gamma_{13}^\prime \tilde{\gamma}_3 \gamma_{32}^\prime \tilde{\gamma}_2 \gamma_{21}^\prime +\gamma_{12}^\prime \tilde{\gamma}_2 \gamma_{23}^\prime \tilde{\gamma}_3 \gamma_{31}^\prime, \label{eq:fourth_order}
\end{equation}
see also Eq.~\eqref{eq:contribution_fifth}. Using the explicit form of $\gamma^\prime_{ij}$ [see Eq.~\eqref{eq:elementary_scatter_process}] allows to rewrite the only relevant contributions as (see Appendix~\ref{app:general_solution})
\begin{equation}
    \sum_{\eta=\uparrow,\downarrow}r_\rS \gamma_\rS \tilde{t}_{\rS\bar{\eta}} \tilde{\gamma}_{\bar{\eta}\bar{\eta}} t_{\bar{\eta} \rS} \gamma_\rS \tilde{t}_{\rS\eta} \tilde{\gamma}_{\eta\eta} t_{\eta \rS} \gamma_\rS \tilde{r}_\rS, \label{eq:current_contribution}
\end{equation}
where $\bar{\eta}$ is the spin opposite to spin $\eta$. {The diagrammatic representation following from the above equation for $\eta=\uparrow$ is shown in Fig.~\ref{fig:lowest_current_contribution}, and the corresponding diagram for $\eta=\downarrow$ is obtained by exchanging red and blue colors.} The remaining contributions to Eq.~\eqref{eq:fourth_order} are discussed in Appendix~\ref{sec:app:diagrams} (see, in particular, Fig.~\ref{fig:fourth_order_Andreev_processes}), where we show that they do not contribute to the supercurrent for the similar reasons discussed previously. 

These processes (see Eq.~\eqref{eq:current_contribution} and Fig.~\ref{fig:lowest_current_contribution}) known as \textit{crossed pair transmission processes} \cite{greinSpinDependentCooperPair2009} effectively transmit two singlet Cooper pairs from the SC as two equal-spin triplet pairs with opposite spin orientation in the sFM. Since two Cooper pairs are transmitted it gives rise to the second harmonic $\sim\sin(2\Delta\chi)$ in the Josephson CPR and we show this explicitly below (see Sec.~\ref{sec:Analytic_approx}). However, such processes should not be understood naively as schematically shown in Fig.~\ref{fig:system_even_harmonics}(b), but in the spirit of Fig.~\ref{fig:lowest_current_contribution}. Specifically, the lowest nonvanishing process involves not two, but four spin-singlet Cooper pairs in the SC, among which three are incoming and one is outgoing; therefore, effectively, two Cooper pairs are transmitted. Here, we add two important remarks. First, such processes are only allowed for nonvanishing incoming coherence amplitudes from the sFM, $\gamma_{\eta\eta}$ and $\tilde{\gamma}_{\eta\eta}$. Therefore a magnetic inhomonegenity at the left SC/sFM interface is demanded. Second, as seen from Fig.~\ref{fig:lowest_current_contribution}, such processes are possible only if both spin-bands in sFM are involved. Therefore, the effect is absent in the case of a half-metallic ferromagnet. 

To conclude this subsection, we comment on the contribution of next order in Eq.~\eqref{eq:Gamma_1}, namely,
\begin{equation}
    \gamma_{13}^\prime \tilde{\gamma}_3 \gamma_{32}^\prime \tilde{\gamma}_2 \gamma_{23}^\prime \tilde{\gamma}_3 \gamma_{31}^\prime,
\end{equation}
for which the corresponding diagrams are shown in Appendix~\ref{sec:app:diagrams} (see Fig.~\ref{fig:sixth_order_Andreev_processes}). The processes in Fig.~\ref{fig:sixth_order_Andreev_processes}(a) and (d) cannot contribute to the current because they lack spin-flip processes. The two remaining diagrams contribute to the supercurrent giving rise to a a second harmonic in the CPR. They are closely related to the lower order process shown in Fig.~\ref{fig:lowest_current_contribution} and the difference is that either $\tilde{r}_\rS$ or $r_\rS$ are replaced by 
\begin{align}
    \tilde{r}_\rS &\rightarrow \tilde{t}_{\downarrow \rS} \tilde{\gamma}_{\downarrow\downarrow} r_\downarrow \gamma_{\downarrow\downarrow} \tilde{t}_{\downarrow \rS}, \\
    r_\rS &\rightarrow t_{\rS \downarrow} \gamma_{\downarrow\downarrow} \tilde{r}_\downarrow \tilde{\gamma}_{\downarrow\downarrow} t_{\downarrow \rS}.
\end{align}
Thus, we obtain that the lowest nonvanishing contribution from Eq.~\eqref{eq:Gamma_1} to the supercurrent is of the fourth order in the transmission amplitudes giving rise to a second harmonic. Finally, the contributions contained in $\Gamma_1^\mathrm{higher}$ give to rise to either a second but more generally to higher even harmonics. However, the latter are typically strongly suppressed, as we show below. For more details on higher-harmonic contributions, see  Appendix~\ref{app:general_solution}.
\subsection{Analytic approximation}\label{sec:Analytic_approx}
So far we provided a purely diagrammatic analysis of the Andreev processes at the right SC/sFM interface. In the following, we consider the case of a tunnel junction which allows to explicitly obtain the second harmonic as the leading contribution to the Josephson CPR. The scattering matrix (see Appendix~\ref{app:scatmat}) in the tunneling limit is characterized by $\abs{t_{\eta \rS}},\abs{t_{\rS \eta}} \ll \abs{r_{\rS\eta}}, \abs{r_\eta}$, $\eta = \uparrow,\downarrow$. Consequently, the superconductor is negligibly influenced by the inverse proximity effect, and we consider as incoming amplitudes in the SC the homogeneous BCS solutions, $\gamma_\rS^\mathrm{hom}$ (see Appendix~\ref{app:solution_hom_S}). Using the notation for the S-matrix as shown in Appendix~\ref{app:scatmat}, the outgoing amplitude in the superconductor [see Eq.~\eqref{eq:current_contribution}] can be evaluated as
\begin{equation}\label{eq:approx:contribution}
    \Gamma_1 \approx -2\gamma_0^3\tilde{\gamma}_{\uparrow\uparrow} \tilde{\gamma}_{\downarrow\downarrow} |t_\uparrow|^2|t_\downarrow|^2 |r_{\rS\uparrow}||r_{\rS\downarrow}|\begin{pmatrix}
        0 & e^{i\phi}\\
        -e^{-i\phi} & 0
    \end{pmatrix},
\end{equation} 
where $\gamma_0$ is given in Eq.~\eqref{eq:homogeneous_gamma_S}, $\tilde{\gamma}_{\eta\eta}$ represents the hole-like incoming amplitudes from the sFM side, and $\phi$ refers to the spin-mixing angle {acquired upon reflection at the right interface}. Note that the latter is small since the insulating layer is not spin active. Applying the particle-hole conjugation operation, we obtain the hole-like outgoing amplitude in the superconductor
\begin{equation}\label{eq:approx:tilde_contribution}
    \tilde{\Gamma}_1 \approx -2 \tilde{\gamma}_0^3\gamma_{\uparrow\uparrow} \gamma_{\downarrow\downarrow} |t_\uparrow|^2|t_\downarrow|^2 |r_{\rS\uparrow}||r_{\rS\downarrow}|\begin{pmatrix}
        0 & e^{-i\phi}\\
        -e^{i\phi} & 0
    \end{pmatrix}.
\end{equation}
Finally, the supercurrent across the right SC/sFM interface for a single trajectory $j$ can be approximated by [see Eqs.~\eqref{eq:j_S}-\eqref{eq:fs_to_gammas}]
\begin{equation}\label{eqn:current_linearized}
    j \approx \frac{N_F v_F}{2} \Tr_2\qty(\Gamma_1 \tilde{\gamma}_\rS^\mathrm{hom}-\tilde{\Gamma}_1\gamma_\rS^\mathrm{hom}), 
\end{equation}
where the trace is performed over spin space only. Inserting Eqs.~\eqref{eq:approx:contribution} and~\eqref{eq:approx:tilde_contribution} into the expression above, we arrive at
\begin{equation}
    j\!\approx\!N_Fv_F|t_\uparrow|^2|t_\downarrow|^2 |r_{\rS\uparrow}||r_{\rS\downarrow}| \gamma_0\tilde{\gamma}_0\qty[\gamma_0^2\tilde{\gamma}_{\uparrow\uparrow}\tilde{\gamma}_{\downarrow\downarrow}\!-\!\tilde{\gamma}_0^2\gamma_{\uparrow\uparrow} \gamma_{\downarrow\downarrow}]\!\cos\phi.
\end{equation}
Since we assume a ferromagnetic insulating layer at the left SC/sFM interface, the $\gamma_{\eta\eta}$ amplitudes are nonzero. In the tunneling limit, these amplitudes can be approximated to the leading order by $\gamma_{\eta\eta} \propto \sin\alpha\, e^{i\chi_L\mp i\varphi}$~\cite{greinSpinDependentCooperPair2009}, where $-(+)$ is for $\eta=\uparrow(\downarrow)$ and $\alpha$ and $\varphi$ are, respectively, the polar and the azimuthal angle characterizing the direction of the exchange field of the ferromagnetic insulator on the left. Keeping in mind that $\gamma_0 \propto e^{i\chi_R}$, we finally obtain
\begingroup
\allowdisplaybreaks
\begin{align}
   j&\propto |t_\uparrow|^2 |t_\downarrow|^2  |r_{\rS\uparrow}||r_{\rS\downarrow}|\sin^2\!\alpha\cos\phi \qty (e^{2i\Delta\chi}\!-\!e^{-2i\Delta\chi})\nonumber\\
    &=2i |t_\uparrow|^2 |t_\downarrow|^2  |r_{\rS\uparrow}||r_{\rS\downarrow}|\sin^2\!\alpha\cos\phi\sin(2\Delta\chi), \label{eq:tunnel_approx}
\end{align}
where $\Delta\chi=\chi_R-\chi_L$ is the superconducting phase difference. Thus, the second harmonic emerges as the leading contribution to the Josephson current in the tunneling limit. Note that the effect is absent in the case of $\alpha=0$. Therefore, the misalignment between the exchange fields of the ferromagnetic insulating layer at the left SC/sFM interface and the metallic ferromagnet is necessary. Remarkably, the simple dependence of $j\propto\sin^2\!\alpha$ proves to be a good approximation to the full numerical solution as well even in the diffusive limit presented in the subsequent section.
\endgroup
\section{Diffusive limit}
In the previous section, we have considered the ballistic limit allowing for an analytic study of the underlying Andreev processes. As we have shown, all lower-order processes cannot contribute to the supercurrent and that the first nonvanishing contribution comes from Andreev processes of fourth order in transmission amplitudes giving rise to the second harmonic in the Josephson CPR. In this section, we numerically confirm the above findings, by studying the same system in the diffusive limit. This regime assumes the electronic mean free path $\ell$ being much smaller than the superconducting coherence length of a ballistic superconductor, i.e., $\ell \ll \xi_\mathrm{c}$. 
Similarly to the previous section, we first outline the theoretical framework, before discussing the results.
\subsection{Theory} \label{sec:theory:diffusive}
As for the ballistic case (see Sec.~\ref{sec:theory:ballistic}), in this subsection, we describe each constituent of the system shown in Fig.~\ref{fig:system_even_harmonics}(a) in the diffusive regime.
\subsubsection{Superconductor} \label{sec:theory:diffusive:SC}
In the diffusive limit, the angular dependence of the quasiclassical GF can be expanded in spherical harmonics up the the linear order, leading to the diffusive Green's function $\hat{G}(\vec{R},E)$~\cite{alexanderTheoryFermiliquidEffects1985}:
\begin{equation}
    \hat{G} = \begin{pmatrix} \mathcal{G} & \mathcal{F} \\ -\tilde{\mathcal{F}} & -\tilde{\mathcal{G}} \end{pmatrix}.
\end{equation}
Each entry of the above matrix is a $2\times2$ matrix itself in spin space, and the standard parametrization is used~\cite{sereneQuasiclassicalApproachSuperfluid1983}: 
\begin{equation}
    \hat{G} = \begin{pmatrix} \mathcal{G}_0 \mathit{1} + \bm{\mathcal{G}}\cdot\bm{\sigma} & (\mathcal{F}_0 \mathit{1}  + \bm{\mathcal{F}}\cdot\bm{\sigma})i\sigma_y, \\
    -i\sigma_y(\tilde{\mathcal{F}}_0 \mathit{1} + \tilde{\bm{\mathcal{F}}}\cdot\bm{\sigma}) & -(\tilde{\mathcal{G}}_0 \mathit{1} + \tilde{\bm{\mathcal{G}}}\cdot\bm{\sigma}^*) \end{pmatrix}.
\end{equation}
Here, $\bm{\sigma}=(\sigma_x,\sigma_y,\sigma_z)^T$ is the Pauli vector in spin space, $\mathit{1} = \mathrm{diag}(1,1)$ is the identity matrix, $\mathcal{F}_0$ and $\bm{\mathcal{F}} = (\mathcal{F}_x,\mathcal{F}_y,\mathcal{F}_z)^T$ are the singlet and triplet pair amplitudes, respectively, and $\mathcal{G}_0$ and $\bm{\mathcal{G}}=(\mathcal{G}_x,\mathcal{G}_y,\mathcal{G}_z)^T$ are the spin-scalar and spin-vector contributions to $\mathcal{G}$, respectively. Here, the particle-hole symmetry relation is defined as $\tilde{\mathcal{Q}}(\vec{R},E) = \mathcal{Q}^\ast(\vec{R},-E^\ast)$. The equation of motion for the GF in the diffusive limit is given by the Usadel equation~\cite{usadelGeneralizedDiffusionEquation1970}
\begin{equation}\label{eq:Usadel}
    - i\hbar \sum_{k,l} D_{kl} \grad_k \qty(\hat{G}\grad_l\hat{G}) +  \comm{E\hat{\tau}_3 - \hat{\Delta}}{\hat{G}} = \hat{0},
\end{equation}
which is supplemented by the normalization condition $\hat{G} = \hat{1}^2$. Here, $D_{kl}$ with $k,l \in \{x,y,z\}$ is the diffusion tensor and the gap matrix $\hat{\Delta}$ for spin-singlet BCS pairing is given by 
\begin{equation}
    \hat{\Delta} = \begin{pmatrix} 0 & \Delta \\ \Delta^* & 0 \end{pmatrix}.
\end{equation}
The pairing potential is determined selfconsistently from the Green's function as 
 \begin{equation}
     \Delta(\vec{R}) = \lim_{E_c \rightarrow \infty} \frac{-\int_{-E_c}^{E_c} \frac{dE}{2} \mathcal{F}_0(\vec{R},E) \tanh{\frac{E}{2k_B T}}}{\ln{\frac{T}{T_c}} + \int_{-E_c}^{E_c}\frac{dE}{2E} \tanh{\frac{E}{2k_B T}}} i\sigma_y, \label{eq:gap}
\end{equation}
where $T_c$ is the superconducting critical temperature. The results presented below are obtained for a quasi-one-dimensional system geometry and isotropic Fermi surfaces implying a diagonal diffusion tensor, i.e, $D_{kl} = D \delta_{kl}$ with $D = v_F^2\tau/3$. Here $\tau$ is the impurity scattering time. Furthermore, all gradients in the previous equations simplify to a partial derivative along the respective axis of the system, which is here assumed to be the $z$-axis.

The formalism presented above has been formulated under the assumption of spin coherence between two spin bands, which is maintained in, e.g., superconductors, normal metals, or weakly spin-polarized magnetic materials. In the latter, the exchange field can be introduced in a usual way by $E\to E-\vec{J}\cdot\bm{\sigma}$, where $\vec{J}=J(\sin\alpha\cos\varphi,\sin\alpha\sin\varphi,\cos\alpha)^T$ denotes the exchange field vector. However, such a procedure cannot be made in strongly spin-polarized materials for which $J\sim E_F$, and the approach outlined above needs to be modified, similarly as in the ballistic case described in the previous section.

\subsubsection{Strongly spin-polarized ferromagnet}

As for the ballistic case [see Sec.~\ref{sec:ballistic:sFM}], in strongly spin-polarized materials, the quantum coherence is maintained only within each spin-band characterized by a diffusive propagator $\breve{G}_{\eta\eta}$, $\eta=\uparrow,\downarrow$. The corresponding Usadel equation reads
\begin{equation}
     - i\hbar D_\eta\pdv{}{z}\qty(\breve{G}_{\eta\eta}\pdv{}{z}\breve{G}_{\eta\eta}) + \comm{E\breve{\tau}_3}{\breve{G}_{\eta\eta}} = \breve{0},
\end{equation}
and it is supplemented with the normalization condition $\breve{G}_{\eta\eta}^2 = \breve{1}$. The strong splitting between the spin-bands enters via the spin-dependent diffusion constant $D_\eta$, and the GF has the following matrix structure in particle-hole space:
\begin{equation}
    \breve{G} = \begin{pmatrix} \mathcal{G}_{\eta\eta} & \mathcal{F}_{\eta\eta} \\ -\tilde{\mathcal{F}}_{\eta\eta} & -\tilde{\mathcal{G}}_{\eta\eta} \end{pmatrix}.
\end{equation}
Similarly to Eqs.~\eqref{eq:G_F} and~\eqref{eq:F_F}, the entries of the above matrix can be parameterized by spin-scalar Riccati amplitudes as
\begin{align}
    \mathcal{F}_{\eta\eta} &= (1-\gamma_{\eta\eta}\tilde{\gamma}_{\eta\eta})^{-1}2\gamma_{\eta\eta}, \\
    \mathcal{G}_{\eta\eta} &= (1-\gamma_{\eta\eta}\tilde{\gamma}_{\eta\eta})^{-1} (1+\gamma_{\eta\eta}\tilde{\gamma}_{\eta\eta}),
\end{align}
while the tilded quantities $(\tilde{\,})$ are obtained from the particle-hole conjugation. Having obtained the Green's functions, the spin-resolved charge current can be computed as 
\begin{equation}
    I_{\eta\eta} = \int_{-\infty}^\infty dE j_{\eta\eta}(E) \tanh(\frac{E}{2k_B T}),
\end{equation}
with the spectral current 
\begin{equation}
    j_{\eta\eta}(E) = \frac{N_{F,\eta} D_\eta}{8} \Re\left[\Tr_2\qty(\breve{\tau}_3 \breve{G}_{\eta\eta} \partial_z \breve{G}_{\eta\eta})\right], \label{eq:spec_curr_spin_resolv}
\end{equation}
where $N_{F,\eta}$ is the spin-resolved density of states at the corresponding Fermi level and $\Tr_2$ here denotes the trace over the particle-hole space only. The total charge current is $2eI_\mathrm{ch}$, and in the following we discuss $I_\mathrm{ch}=I_\upup+I_\dodo$. More details on the implementation of the above procedure can be found in Refs.~\cite{ouassouTripletCooperPairs2017,schulzQuantumGeometricSpinCharge2025,schulzTheoryQuantumgeometricCharge2025}.

\subsubsection{Boundary conditions}

The insulating layers at the SC/sFM interfaces enter the theory as the effective boundary conditions, which have been derived in \cite{eschrigGeneralBoundaryConditions2015} and further developed in \cite{schulzTheoryQuantumgeometricCharge2025}. For the insulating layer on the right side of the sFM, we assume a simple box-shaped potential with the same height and width for both spin projections. On the other hand, the left insulating layer is modelled by box-shaped potential with spin-dependent barrier widths as discussed in \cite{schulzTheoryQuantumgeometricCharge2025}. As mentioned earlier, the exchange field of the ferromagnetic insulator points in an arbitrary direction with respect to that of the metallic ferromagnet, $\vec{J}_\mathrm{FI}=J_\mathrm{FI}(\sin\alpha\cos\varphi,\sin\alpha\sin\varphi,\cos\alpha)^T$. Both ferromagnetic regions are modeled by parabolic bands with spin-splitting, $\xi_\eta(\vec{p}) = \frac{\vec{p}^2}{2m} - E_F + V \mp J/2$, $\eta=\uparrow,\downarrow$, where $E_F$ is the Fermi energy of the superconductor, $V$ is a potential bias shifting both spin-bands, and $J$ is the exchange field strength. 

\subsection{Results} \label{sec:diffusive:results}
In the following, we present the numerical results for the Josephson current-phase relation, $I_\mathrm{ch}(\Delta\chi)$ with $\Delta\chi=\chi_2-\chi_1$, obtained self-consistently as described in the preceding subsection. We consider a metallic ferromagnet of thickness $L_\mathrm{sFM}=1\xi_\mathrm{d}$ coupled to two superconducting electrodes of thicknesses $L_\mathrm{SC}=5\xi_\mathrm{d}$ each, and which are treated self-consistently. To ensure the current conservation, the superconducting leads are attached to two bulk BCS superconductors at the outer interfaces characterized by the superconducting phases $\chi_{1/2}$, which enter the CPR shown in figures below. As the existence of the effect presented below is insensitive to the exact values of the exchange fields, we assume it in all ferromagnetic regions the same, $J = 0.4 E_F$. In addition, in the metallic ferromagnet, we set $V=0$, whereas in the left ferromagnetic insulator the potential $V$ shifts the bands such that both are insulating, i.e., $V_\mathrm{FI}^\uparrow = 1.1 E_F$ and $V_\mathrm{FI}^\downarrow = 1.5 E_F$. The right insulating layer is not spin-active and its spin-bands are both assumed to be situated at $V_\mathrm{I} = 1.1 E_F$. The temperature of the system is fixed to $0.1 T_c$. Finally, the Josephson current is scaled in units of $I_0 = e^2R_N / k_B T_c$, where $R_N^{-1} = \sigma$ is the normal-state conductivity of the sFM.

\begin{figure}[t!]
    \centering
    \includegraphics[width=\linewidth]{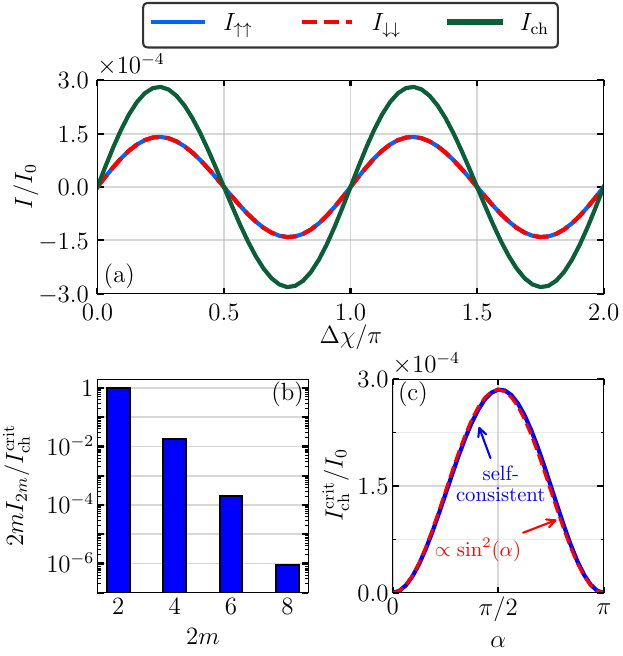}
    \caption{(a) The current-phase relations for the spin-resolved and the total charge current for $\alpha = \pi/2$ and the same insulators' widths $d_L = d_R =0.6\lambda_F/2\pi$. Panel (b) shows the Fourier components corresponding to the CPR shown in panel (a). Panel (c) displays the critical charge current $I_\mathrm{ch}^\mathrm{crit}$ as a function of the tilt of the left barrier's exchange field $\alpha$.}
    \label{fig:CPR_FT_crit_curr_alpha}
\end{figure}

First, we consider the Josephson current-phase relation for the case of $\alpha = \pi/2$, which maximizes the effect, shown in Fig.~\ref{fig:CPR_FT_crit_curr_alpha}(a). Apparently, the CPR is dominated by a second harmonic with higher harmonics being significantly suppressed. In addition, the spin-resolved currents $I_{\uparrow\uparrow / \downarrow\downarrow}$ coincide and contribute equally to the total charge current across the sFM. To quantify the contributions from different harmonics to the CPR, we perform the following expansion~\cite{schulzQuantumGeometricSpinCharge2025,schulzTheoryQuantumgeometricCharge2025}:
\begin{equation}
    I_\mathrm{ch} = \sum_{m=1}^\infty m  I_{m} \sin(m\Delta\chi),
\end{equation} 
where $m$-th term corresponds to the transmission of $m$ Cooper pairs across the junction. The first eight amplitudes of the above expansion are shown in Fig.~\ref{fig:CPR_FT_crit_curr_alpha}(b). We obtain that even harmonic contributions for $m\geq 4$ are roughly exponentially suppressed with $m$, whereas odd harmonic contributions are completely absent within our computation accuracy. Our numerical findings are in good agreement with the expectations following from the analytic analysis performed in Sec.~\ref{sec:ballistic:results}. Thus, the charge supercurrent in the system is well described by $I_\mathrm{ch} \approx 2 I_2 \sin(2\Delta\chi)$.

Next, we consider the functional dependence of the critical charge current $I_\mathrm{ch}^\mathrm{crit}$ on different system parameters. In Fig.~\ref{fig:CPR_FT_crit_curr_alpha}(c), we show the critical current $I_\mathrm{ch}^\mathrm{crit}$ as a function of the misalignment angle $\alpha$. The maximum critical current is achieved for $\alpha=\pi/2$ and the functional dependence is well approximated by $I_\mathrm{ch}^\mathrm{crit}\propto\sin^2\!\alpha$, which is in a good agreement with Eq.~\eqref{eq:tunnel_approx}. This dependence can be explained by considering a phenomenological model for the triplet pair amplitudes in the ferromagnetic insulator and the metallic ferromagnet. The spin rotation from one quantization axis to the other is given by
\begin{equation}\label{eq:spin_rotation}
    \begin{pmatrix} \uparrow \\ \downarrow \end{pmatrix}_ {\vec{J}_\mathrm{FI}} = \begin{pmatrix} \cos{\frac{\alpha}{2}} & \sin{\frac{\alpha}{2}} \\ - \sin{\frac{\alpha}{2}} & \cos{\frac{\alpha}{2}} \end{pmatrix} \begin{pmatrix} \uparrow \\ \downarrow \end{pmatrix}_{\vec{J}_\mathrm{sFM}},
\end{equation}
where we assumed that $\varphi = 0$ [see Sec.~\ref{sec:system} and Fig.~\ref{fig:system_even_harmonics}(a)]. Consequently, the mixed-spin triplet transforms as
\begin{equation} \label{eq:triplet_rotation}
    \begin{split}
        \qty(\uparrow\downarrow + \downarrow\uparrow)_{\vec{J}_\mathrm{FI}} = &-\sin\alpha \qty[\qty(\uparrow\uparrow)_{\vec{J}_\mathrm{sFM}} -\qty(\downarrow\downarrow)_{\vec{J}_\mathrm{sFM}}] 
        \\ &+ \cos \alpha \qty(\uparrow \downarrow + \downarrow\uparrow)_{\vec{J}_\mathrm{sFM}}.
    \end{split}
\end{equation} 
Thus, a process involving two equal-spin triplet pairs scales with $\sin^2\!\alpha$ as the amplitude for each of them scales as $\sin\alpha$. As a result, the critical current scales similarly [see Fig.~\ref{fig:CPR_FT_crit_curr_alpha}(c)] as the supercurrent is dominantly mediated by two equal-spin pairs with opposite spin orientations (see also Sec.~\ref{sec:ballistic:results}). For the case of collinear arrangements of the exchange fields, i.e., $\alpha = 0,\pi$, there is no supercurrent irrespective of the superconducting phase difference due to the absence of the equal-spin triplet correlations.

\begin{figure}
    \centering
    \includegraphics[width=0.9\linewidth]{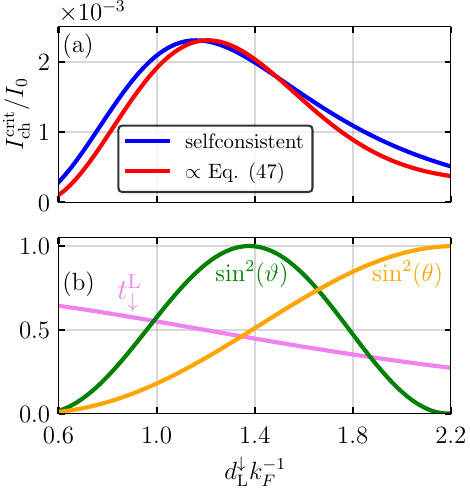}
    \caption{(a) The critical current as a function of the barrier width for the spin-$\downarrow$ channel of the spin-active (left) insulator, which modulates the spin-mixing angles. The blue curve denotes the fully self-consistent solution, whereas the red line denotes the approximate scaling from Eq.~\eqref{eq:approx_sma}. Panel (b) shows the functional dependence of the constituents of Eq.~\eqref{eq:approx_sma} on the barrier width of the spin-$\downarrow$ channel. We do not show $t^\mathrm{L}_\uparrow$, as it is almost constant as a function of $d_\mathrm{L}^\downarrow$. Other parameters are the same as for Fig.~\ref{fig:CPR_FT_crit_curr_alpha}(a).}
    \label{fig:Crit_curr_sma}
\end{figure}

{As shown above, the existence of equal-spin triplet correlations, and consequently the overall effect, crucially depends on the mixed-spin triplet correlations. In strongly spin-polarized materials, they are nonvanishing only in an atomically thin layer close to the SC/sFM interface. Considering the ferromagnetic bilyer described previously [see Fig.~\ref{fig:system_even_harmonics}(a)], we completely neglect them in the sFM, which extends over mesoscopic length scales. However, we allow for them in the atomically thin ferromagnetic insulator located between the SC and sFM, and the singlet-triplet conversion in this region is enabled by the spin-mixing mechanism.} Namely, upon a reflection at the spin-active region spin-$\uparrow$ and spin-$\downarrow$ electrons in the superconductor acquire different phases determined by the so-called spin-mixing angle, $\vartheta$, yielding the following transformation
\cite{eschrigSpinpolarizedSupercurrentsSpintronics2011}:
\begin{equation}
    \qty(\uparrow\downarrow - \downarrow\uparrow) \!\rightarrow\!\cos\vartheta \qty(\uparrow\downarrow - \downarrow\uparrow) + i \sin\vartheta\qty(\uparrow\downarrow + \downarrow\uparrow). \label{eq:spin_mixing}
\end{equation}
Thus, we see that the amplitude of the mixed-spin triplet pairs scales as $\sin{\vartheta}$, and it is expected that the amplitude of each equal-spin triplet pair scales similarly [see Eq.~\eqref{eq:spin_rotation}]. The spin-mixing angle can be increased by increasing the mismatch in the spin-resolved barrier widths in our model for the ferromagnetic insulator (for more details, see Sec.~III~B in Ref. \cite{schulzTheoryQuantumgeometricCharge2025}). In the following, we assume that $d_\mathrm{L}^\uparrow = 0.6 \lambda_F / 2\pi$ and vary $d_\mathrm{L}^\downarrow$. Note that the role of increasing $d_\mathrm{L}^\downarrow$ is not straightforward. On one hand it increases the spin-mixing angle enhancing the effect, however, on the other hand the transmission of spin-$\downarrow$ channel is decreased, suppressing the effect. Therefore, there is an interplay between these two competing effect that should be taken into account, as it is clearly visible in Fig.~\ref{fig:Crit_curr_sma}. Before we comment on it, let us introduce another spin-mixing angle and explain its role.

In the literature, spin-mixing is often motivated by considering the reflection of a singlet Cooper pair at an spin-active interface, as in Eq.~\eqref{eq:spin_mixing}. However, spin-mixing also occurs during a transmission process as well. Therefore, we expect that the amplitude of the equal-spin correlations and, consequently, the critical current is not only a function of the spin-mixing angle due to reflection, $\vartheta$, but also a function of the spin-mixing angle acquired during transmission, $\theta$. The definitions of the these spin-mixing angles are provided in Appendix~\ref{app:scatmat}.
If a Cooper pair injected from the superconductor undergoes a normal transmission process, the spin-mixing implies the presence of mixed-spin triplet correlations in the ferromagnetic insulating layer, whose amplitude scales as $\sin\theta$ [formula is analog to Eq.~\eqref{eq:spin_mixing}].

As already mentioned, the other process leading to spin-mixing is when a singlet Cooper pair is reflected at the interface. The mixed-spin triplet contribution to the reflected pair is proportional to $\sin\vartheta$ [see Eq.~\eqref{eq:spin_mixing}]. Due to impurity scattering such a reflected pair, containing mixed-spin triplet correlations, can scatter back to the interface. As for the singlet Cooper pair [see Eq.~\eqref{eq:spin_mixing}], the mixed-spin triplet correlations of the incoming pair acquire a spin-mixing phase upon transmission. The amplitude of the mixed-spin triplets created from this process scales as $\sin\vartheta \cos\theta$. Furthermore, each of the processes involve the transmission of one spin-$\uparrow$ and one spin-$\downarrow$ particle across the interface, which leads to an additional scaling factor of $t_\uparrow^\mathrm{L} t^\mathrm{L}_\downarrow$ (L denoting scattering matrix elements for the left interface). Finally, the analytic discussion from Sec.~\ref{sec:ballistic:results} yields that the leading order process involves the simultaneous transmission of one $\upup$ and one $\dodo$ pair. Therefore, we must multiply the amplitudes of $\upup$ and $\dodo$ pairs to obtain a approximate scaling of the critical current which follows as:
\begin{equation} \label{eq:approx_sma}
    I^\mathrm{crit}_\mathrm{ch,approx}(\vartheta,\theta) \propto \qty(t_\uparrow^\mathrm{L})^2 \qty(t_\downarrow^\mathrm{L})^2 \qty[\sin\vartheta\cos\theta + \sin\theta]^2.
\end{equation}
In Fig.~\ref{fig:Crit_curr_sma}(a), we compare the selfconsitently obtained critical current as a function $d_\mathrm{L}^\downarrow$ (blue) with the approximate scaling from Eq.~\eqref{eq:approx_sma} (red), obtaining qualitative agreement. The deviations can be caused by other types of processes for which we do not account, such as spin-flip transmission processes or generally higher order processes. However, the simple phenomenological picture already gives a good approximation. Note that a nonmonotonic behavior of the $I_\mathrm{ch}^\mathrm{crit}(d_\mathrm{L}^\downarrow)$ is caused by the interplay between the spin-mixing angles and the transmission of spin channels, as mentioned earlier. Here, we quantify it. As shown in Fig.~\ref{fig:Crit_curr_sma}(b), small $d_\mathrm{L}^\downarrow$ reduces drastically the spin-mixing angles important for the creation of mixed-spin triplets in the ferromagnetic insulator making the overall effect weak. Increasing its value, the spin mixing angles increase reaching the critical value of $\pi/2$. However, maximum critical current is reached below this value due to the reduced transmission. Further increase of $d_\mathrm{L}^\downarrow$ suppresses the critical current due to the reduction of the transmission which proves to be crucial in this regime. 

\begin{figure}[t!]
    \centering
    \includegraphics[width=0.9\linewidth]{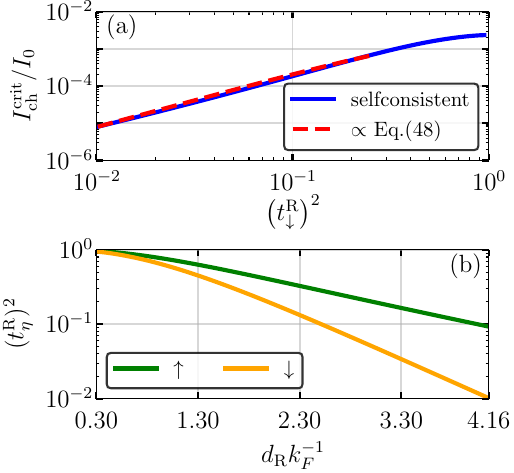}
    \caption{(a) The critical current as a function of the transmission amplitude for the $\downarrow$-channel of the right non spin-active insulator. The solid blue line denotes the self-consistently obtained values and the red dashed line shows the approximate expression obtained in the tunneling limit [see Eq.~\eqref{eq:tunnel_approx}]. Since the approximation in Eq.~\eqref{eq:tunnel_approx} lacks appropriate units we matched it to the lowest transmission probability considered for the self-consistent solution. Panel (b) shows the functional dependence of the transmission probabilities for spin-$\uparrow$ (green) and spin-$\downarrow$ (orange) on the barrier width of the right insulating layer. Left barrier width for the spin-$\downarrow$ channel is $d^\downarrow_\mathrm{L} k_F^{-1} = 1.24$ and other parameters are the same as for Fig.~\ref{fig:CPR_FT_crit_curr_alpha}(a).}
    \label{fig:Crit_curr_right_transm}
\end{figure}

Finally, we consider the critical charge current as a function of the transmission probability of the right interface, determined by the thickness of the insulating layer. The functional dependence of the transmission probabilities on the spin-degenerate barrier width $d_R$ is shown in Fig.~\ref{fig:Crit_curr_right_transm}(b). In the spirit of the analytic consideration of Sec.~\ref{sec:Analytic_approx} [see Eq.~\eqref{eq:tunnel_approx}], for low transparencies we expect the following approximate behavior of the critical current:
\begin{equation}
    I_\mathrm{ch,approx}^\mathrm{crit} \propto \abs{r_{\rS\uparrow}^\mathrm{R}} \abs{r_{\rS\downarrow}^\mathrm{R}} \abs{t_\uparrow^\mathrm{R}}^2 \abs{t_\downarrow^\mathrm{R}}^2.    
\end{equation}
The critical current as a function of the transmission probability across the right insulating layer is shown in Fig.~\ref{fig:Crit_curr_right_transm}(a). The blue line denotes the full selfconsistent solution, while the red dashed line shows the approximate solution given above. Remarkably, the agreement is good even beyond the tunneling limit, namely, even for $(t_\downarrow^\mathrm{R})^2>0.1$. However, for highly transparent cases the approximation is not valid anymore since higher-order Andreev processes need to be taken into account. Finally, as mentioned, the transparency is controlled by the thickness of the insulator, and Fig.~\ref{fig:Crit_curr_right_transm}(b) shows this dependence for the spin-$\uparrow$ (green) and the spin-$\downarrow$ channel (orange). 

\section{Conclusion}
In summary, we have provided a theoretical study of the Josephson current across a strongly spin-polarized metallic ferromagnet (sFM) coupled to two BCS superconductors (SC) via a thin ferromagnetic insulating layer (FI) on the left, and a thin non spin-active (I) insulating layer on the right, both in the clean and the diffusive limit. A misalignment between the exchange fields of the FI and the sFM allows for the creation of equal-spin triplet pairs which can then enter the latter. Applying a finite superconducting phase difference between the superconductors adjacent to the sFM yields a purely even-harmonic Josephson supercurrent. We have shown that the nature of these contributions is long-ranged as the current is fully mediated by $\upup$ and $\dodo$ triplet correlations due to the strong spin polarization of the sFM. Using a diagrammatic technique suited for ballistic propagators, we have provided a full microscopic picture of the effect showing that the lowest order process that gives rise to a second harmonic amplitude in the CPR involves four Cooper pairs in the SC, among which three are incoming and one is outgoing. This process leads to an effective transfer of the net charge of $4e$, which manifests itself as a second harmonic in the Josephson CPR. In addition, higher order processes have been also considered showing that only an even number of pairs can be transferred across the junction. Finally, we have treated the system in the diffusive limit numerically confirming the previous above.

\section*{Acknowledgement}
We acknowledge Alexander I. Buzdin, Manuel Houzet, Tomas Löfwander, and Zoran Radović for valuable discussions. NLS and ME acknowledge funding by the Deutsche Forschungsgemeinschaft (DFG, German Research
Foundation) under project number 530670387. The computations were enabled by resources provided by the University Computer Centre of the University of Greifswald.
\appendix

\section{Bulk homogeneous BCS superconductor} \label{app:solution_hom_S}
For the case of a bulk BCS superconductor the coherence amplitude solving Eqs.~\eqref{eq:eilenberger} and \eqref{eq:Usadel} is
\begin{equation}\label{eq:homogeneous_gamma_S}
    \gamma_\rS^\mathrm{hom}(E) = \frac{-\Delta}{E +i\sqrt{|\Delta|^2-E^2}} i\sigma_2 = \gamma_0(E)i\sigma_2,
\end{equation}
and tilded quantities can be obtained by the particle-hole conjugation operation. Note that the retarded GF assumes the Dynes shift $E \to E +i\eta$~\cite{dynesDirectMeasurementQuasiparticleLifetime1978,dynesTunnelingStudySuperconductivity1984}.

\section{General structure of S-matrices} \label{app:scatmat}
The scattering matrix is calculated from a wave function matching technique, arriving at the following matrix form in channel space
\begin{equation}
    \bm{S} = \begin{pmatrix} r_\rS & t_{\rS\uparrow} & t_{\rS\downarrow} \\ t_{\uparrow \rS} & r_\uparrow & r_{\downarrow\uparrow} \\ t_{\downarrow \rS} & r_{\uparrow\downarrow} & r_\downarrow \end{pmatrix},
\end{equation}
Here $r_\rS$ is a $2\times2$ matrix in spin space and $t_{\rS\eta}$ are column vectors, $\eta=\uparrow,\downarrow$. 
\begin{equation}
    r_\rS = \begin{pmatrix} r_{\rS \uparrow} & r_{\rS \uparrow\downarrow} \\ r_{\rS\downarrow\uparrow} & r_{\rS \downarrow} \end{pmatrix}, \quad t_{\rS\eta} = \begin{pmatrix} t_\eta \\ t_{\eta\bar\eta} \end{pmatrix},
\end{equation}
and $t_{\eta\rS} = t_{\rS\eta}^T$ is a row vector. The spin-mixing angle upon reflection is defined as ${\vartheta = \vartheta_\uparrow - \vartheta_\downarrow}$, where $r_{\rS\eta} = \abs{r_{\rS\eta}} e^{i\vartheta_\eta}$. Analogously, the spin-mixing acquired upon transmission is defined as ${\theta = \theta_\uparrow - \theta_\downarrow}$, where $t_\eta = \abs{t_\eta} e^{i\theta_\eta}$. The misalignment between the exchange fields of the ferromagnetic insulator on the left and the metallic ferromagnet results into all the entries in the above matrix being nonzero. However, as the right interface is nonmagnetic, spin-flip processes are absent and the matrix is diagonal in spin space,
\begin{equation}
    \bm{S}^\mathrm{R} = \begin{pmatrix} r_{\rS\uparrow}^\mathrm{R} & 0 & t_{\uparrow}^\mathrm{R} & 0 \\ 0 & r_{\rS\downarrow}^\mathrm{R} & 0 & t_{\downarrow}^\mathrm{R} \\ t_{\uparrow}^\mathrm{R} & 0 & r_\uparrow^\mathrm{R} & 0 \\ 0& t_{\downarrow}^\mathrm{R} & 0 & r_\downarrow^\mathrm{R} \end{pmatrix}.
\end{equation}
For an ease of notation we omit the superscript $\ldots^\mathrm{R}$ in Sections~\ref{sec:ballistic:results} and \ref{sec:Analytic_approx} and in Appendices~\ref{app:general_solution} and \ref{sec:app:diagrams}. 
\section{General solutions to the boundary conditions} \label{app:general_solution}
In this appendix, we provide a more detailed discussion of the boundary conditions, introduced in Sec.~\ref{sec:ballistic:bc} of the main text. Following Ref.~\cite{eschrigScatteringProblemNonequilibrium2009}, we can calculate the outgoing amplitudes as
\begingroup
\allowdisplaybreaks
\begin{subequations}
\begin{align}
    \Gamma_1 &= \gamma_{11}^\prime + \Gamma_{1\leftarrow 2}\, \tilde{\gamma}_2\, \gamma_{21}^\prime + \Gamma_{1\leftarrow 3}\, \tilde{\gamma}_3\, \gamma_{31}^\prime, \\
    \Gamma_2 &= \gamma_{22}^\prime + \Gamma_{2\leftarrow 1}\, \tilde{\gamma}_1\, \gamma_{12}^\prime + \Gamma_{2\leftarrow 3}\, \tilde{\gamma}_3\, \gamma_{32}^\prime, \\
    \Gamma_3 &= \gamma_{33}^\prime + \Gamma_{3\leftarrow 1}\, \tilde{\gamma}_1\, \gamma_{13}^\prime + \Gamma_{3\leftarrow 2}\, \tilde{\gamma}_2\, \gamma_{23}^\prime,
\end{align}
\end{subequations}
\endgroup
where the auxiliary amplitude $\Gamma_{i \leftarrow j}$ for $i,j\in\{1,2,3\}$ and $i\neq j$ are defined as
\begingroup
\allowdisplaybreaks
\begin{subequations}
\begin{align}
    \Gamma_{1\leftarrow 2} &= \qty[\gamma_{12}^\prime + \gamma_{13}^\prime N_3 \tilde{\gamma}_3 \gamma_{32}^\prime] N_2 N_{23}, \label{eq:Gamma_1_2} \\
    \Gamma_{1\leftarrow 3} &= \qty[\gamma_{13}^\prime + \Gamma_{1\leftarrow 2} \tilde{\gamma}_2 \gamma_{23}^\prime] N_3, \label{eq:Gamma_1_3} \\
    \Gamma_{2\leftarrow 1} &= \qty[\gamma_{21}^\prime + \gamma_{23}^\prime N_3 \tilde{\gamma}_3 \gamma_{31}^\prime] N_1 N_{13}, \\
    \Gamma_{2\leftarrow 3} &= \qty[\gamma_{23}^\prime + \Gamma_{2\leftarrow 1} \tilde{\gamma}_1 \gamma_{13}^\prime] N_3, \\
    \Gamma_{3\leftarrow 1} &= \qty[\gamma_{31}^\prime + \gamma_{32}^\prime N_2 \tilde{\gamma}_2 \gamma_{21}^\prime] N_1 N_{12}, \\
    \Gamma_{3\leftarrow 2} &= \qty[\gamma_{32}^\prime + \Gamma_{3\leftarrow 1} \tilde{\gamma}_1 \gamma_{12}^\prime] N_2.
\end{align}
\end{subequations}
In the above expressions $N_{i}$ and $N_{ij}$ are defined as
\begin{align}
    N_i &= \qty(1-\tilde{\gamma}_i\gamma_{ii}^\prime)^{-1}, \\
    N_{ij} &= \qty(1-\tilde{\gamma}_i \gamma_{ij}^\prime N_j \tilde{\gamma}_j \gamma_{ji}^\prime N_i)^{-1}.
\end{align}
\endgroup
Since $\abs{\gamma_i},\abs{\tilde{\gamma}_i},\abs{\gamma_{ij}^\prime} < 1, \forall i,j$, the expressions above can be expanded into geometric series as 
\begin{align}
    N_i &= 1 + \sum_{k=1}^\infty (\tilde{\gamma}_i \gamma_{ii}^\prime)^k \equiv 1 + n_i, \label{eq:app:n_i} \\
    N_{ij} &= 1 + \sum_{k=1}^\infty \left\{ \qty(\tilde{\gamma}_i \gamma_{ij}^\prime \tilde{\gamma}_j \gamma_{ji}^\prime) + \qty(\tilde{\gamma}_i \gamma_{ij}^\prime n_j \tilde{\gamma}_j \gamma_{ji}^\prime) \right. \nonumber \\
    &\left. \qquad + \qty(\tilde{\gamma}_i \gamma_{ij}^\prime \tilde{\gamma}_j \gamma_{ji}^\prime n_i) + \qty(\tilde{\gamma}_i \gamma_{ij}^\prime n_j \tilde{\gamma}_j \gamma_{ji}^\prime n_i) \right\}^k \nonumber\\
    &\equiv 1 + n_{ij}. \label{eq:app:n_ij_full}
\end{align}
Since we are particularly interested in the solution inside the superconductor (see Sec.~\ref{sec:ballistic:results}), we focus on $\Gamma_1$. Using Eqs.~\eqref{eq:app:n_i} and \eqref{eq:app:n_ij_full} allows to rewrite Eqs.~\eqref{eq:Gamma_1_2} and \eqref{eq:Gamma_1_3} as follows:
\begin{widetext}
\begingroup
\allowdisplaybreaks
\begin{align}
    \begin{split}
        \Gamma_{1\leftarrow 2} &= \gamma_{12}^\prime +  \gamma_{13}^\prime \tilde{\gamma}_3 \gamma_{32}^\prime + \gamma_{13}^\prime n_3 \tilde{\gamma}_3 \gamma_{32}^\prime + \qty[\gamma_{12}^\prime + \gamma_{13}^\prime \tilde{\gamma}_3 \gamma_{32}^\prime + \gamma_{13}^\prime n_3 \tilde{\gamma}_3 \gamma_{32}^\prime] \cdot \qty[n_2 + n_{23} + n_2 n_{23}]
    \end{split} \\
    \begin{split}
        \Gamma_{1\leftarrow 3} &= \gamma_{13}^\prime + \gamma_{12}^\prime \tilde{\gamma}_2 \gamma_{23}^\prime + \gamma_{13}^\prime \tilde{\gamma}_3 \gamma_{32}^\prime \tilde{\gamma}_2 \gamma_{23}^\prime + \gamma_{13}^\prime n_3 \tilde{\gamma}_3 \gamma_{32}^\prime \tilde{\gamma}_2 \gamma_{23}^\prime + \qty[\gamma_{12}^\prime + \gamma_{13}^\prime \tilde{\gamma}_3 \gamma_{32}^\prime + \gamma_{13}^\prime n_3 \tilde{\gamma}_3 \gamma_{32}^\prime] \cdot \qty[n_2 + n_{23} + n_2 n_{23}] \cdot \tilde{\gamma}_2 \gamma_{23}^\prime \\
        & + \qty[\gamma_{13}^\prime + \gamma_{12}^\prime \tilde{\gamma}_2 \gamma_{23}^\prime + \gamma_{13}^\prime \tilde{\gamma}_3 \gamma_{32}^\prime \tilde{\gamma}_2 \gamma_{23}^\prime + \gamma_{13}^\prime n_3 \tilde{\gamma}_3 \gamma_{32}^\prime \tilde{\gamma}_2 \gamma_{23}^\prime] \cdot n_3 \\
        & + \qty[\gamma_{12}^\prime + \gamma_{13}^\prime \tilde{\gamma}_3 \gamma_{32}^\prime + \gamma_{13}^\prime n_3 \tilde{\gamma}_3 \gamma_{32}^\prime] \cdot \qty[n_2 + n_{23} + n_2 n_{23}] \cdot \tilde{\gamma}_2 \gamma_{23}^\prime \cdot n_3
    \end{split}
\end{align}
\endgroup
Consequently, the outgoing propagator $\Gamma_1$ can be written as
\begin{equation}
    \begin{split}
        \Gamma_1 &= \gamma_{11}^\prime + \red{\gamma_{12}^\prime \tilde{\gamma}_2 \gamma_{21}^\prime} + \orange{\gamma_{13}^\prime \tilde{\gamma}_3 \gamma_{31}^\prime} + \lila{\gamma_{13}^\prime \tilde{\gamma}_3 \gamma_{32}^\prime \tilde{\gamma}_2 \gamma_{21}^\prime} + \green{\gamma_{12}^\prime \tilde{\gamma}_2 \gamma_{23}^\prime \tilde{\gamma}_3 \gamma_{31}^\prime} + \blue{\gamma_{13}^\prime \tilde{\gamma}_3 \gamma_{32}^\prime \tilde{\gamma}_2 \gamma_{23}^\prime \tilde{\gamma}_3 \gamma_{31}^\prime} + \Gamma^\mathrm{higher}_1. \label{eq:app:Gamma_1} \\
    \end{split}
\end{equation}
Here $\Gamma_1^\mathrm{higher}$ contains contributions of higher orders which all are related to one of the contributions designated in color in the equation above. In other words,
\begin{equation}
    \begin{split}
        \Gamma_1^\mathrm{higher} &= \red{\gamma_{12}^\prime} \cdot  \qty[n_2 + n_{23} + n_2 n_{23}] \cdot \red{\tilde{\gamma}_2 \gamma_{21}^\prime} + \orange{\gamma_{13}^\prime} \cdot n_3 \cdot \orange{\tilde{\gamma}_3 \gamma_{31}^\prime} + \lila{\gamma_{13}^\prime} \cdot \qty[1 + n_3] \cdot \lila{\tilde{\gamma}_3 \gamma_{32}^\prime} \cdot \qty[ n_2 + n_{23} + n_2 n_{23}] \cdot \lila{\tilde{\gamma}_2 \gamma_{21}^\prime} \\
        &\quad + \lila{\gamma_{13}^\prime} \cdot n_3 \cdot \lila{\tilde{\gamma}_3 \gamma_{32}^\prime \tilde{\gamma}_2 \gamma_{21}^\prime} + \green{\gamma_{12}^\prime}  \cdot \qty[ n_2 + n_{23} + n_2 n_{23}] \cdot \green{\tilde{\gamma}_2 \gamma_{23}^\prime} \cdot \qty[ 1 + n_3] \cdot \green{\tilde{\gamma}_3 \gamma_{31}^\prime} \\
        &\quad + \green{\gamma_{12}^\prime \tilde{\gamma}_2 \gamma_{23}^\prime} \cdot n_3 \cdot \green{\tilde{\gamma}_3 \gamma_{31}^\prime} + \blue{\gamma_{13}^\prime} \cdot \qty[ 1 + n_3] \cdot \blue{\tilde{\gamma}_3 \gamma_{32}^\prime} \cdot \qty[1 + n_2 + n_{23} + n_2 n_{23}] \cdot \blue{\tilde{\gamma}_2 \gamma_{23}^\prime} \cdot n_3 \cdot \blue{\tilde{\gamma}_3 \gamma_{31}^\prime} \\
        &\quad + \blue{\gamma_{13}^\prime} \cdot \qty[ 1 + n_3] \cdot \blue{\tilde{\gamma}_3 \gamma_{32}^\prime} \cdot \qty[n_2 + n_{23} + n_2 n_{23}] \cdot \blue{\tilde{\gamma}_2 \gamma_{23}^\prime \tilde{\gamma}_3 \gamma_{31}^\prime} + \blue{\gamma_{13}^\prime} \cdot n_3 \cdot \blue{\tilde{\gamma}_3 \gamma_{32}^\prime \tilde{\gamma}_2 \gamma_{23}^\prime \tilde{\gamma}_3 \gamma_{31}^\prime}.
    \end{split} \label{eq:app:Gamma_higher}
\end{equation}
\end{widetext}
For ease of comparison, we colored the contributions contained in $\Gamma_1^\mathrm{higher}$ in the same color as the underlying contribution in Eq.~\eqref{eq:app:Gamma_1}. By comparing Eqs.~\eqref{eq:app:Gamma_1} and \eqref{eq:app:Gamma_higher}, it is evident that the difference is only given by additional $n_2$, $n_3$, and $n_{23}$ terms. In the following, we consider the corresponding diagrams, showing that also these yield, if any, only even harmonic contributions to the current.

\begin{figure}[t!]
    \centering
    \includegraphics[width=\linewidth]{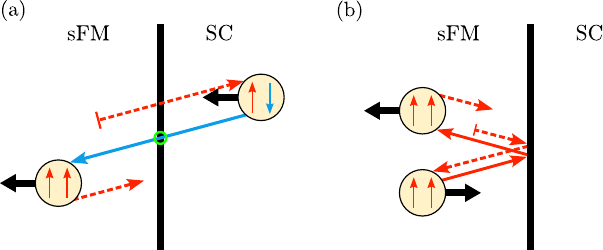}
    \caption{The Andreev processes given by Eq.~\eqref{eq:app:normalisation_lowest_order} for $i=2$, i.e. for spin-up. Panel (a) shows the contribution $\tilde{\gamma}_2 \bm{S}_{21} \gamma_1 \tilde{\bm{S}}_{12} \equiv \tilde{\gamma}_\upup t_{\uparrow\rS} \gamma_\rS \tilde{t}_{\rS\uparrow}$ and panel (b) the term $\tilde{\gamma}_2 \bm{S}_{22} \gamma_2 \tilde{\bm{S}}_{22} \equiv \tilde{\gamma}_\upup r_\uparrow \gamma_\upup \tilde{r}_\uparrow $. The diagrammatic notation is the same as for Fig.~\ref{fig:lower_order_contributions}. The green circle in panel (a) denotes the missing spin-flip process which would be needed for a contribution to the charge current. The diagrams for spin-$\downarrow$ ($i=3$) follow immediatley by exchanging the red and blue colors.}
    \label{fig:app:normalisation_diagrams}
\end{figure}

To diagrammatically represent Eq.~\eqref{eq:app:n_i}, we first consider the lowest order contributions of 
\begin{equation}
    n_i = \tilde{\gamma}_i \gamma_{ii}^\prime + \tilde{\gamma}_i \gamma_{ii}^\prime \tilde{\gamma}_i \gamma_{ii}^\prime + \ldots\ ,
\end{equation}
which are given by 
\begin{equation} \label{eq:app:normalisation_lowest_order}
    \tilde{\gamma}_i \gamma_{ii}^\prime = \tilde{\gamma}_i \bm{S}_{i1} \gamma_1 \tilde{\bm{S}}_{1i} + \tilde{\gamma}_i \bm{S}_{ii} \gamma_i \tilde{\bm{S}}_{ii},
\end{equation}
for $i=2,3$. The Andreev processes following from Eq.~\eqref{eq:app:normalisation_lowest_order} are shown in Fig.~\ref{fig:app:normalisation_diagrams}. Please note that these diagrams are not closed and that the incoming and the outgoing hole have the same spin. This is because they are meant to be inserted into another diagram [see Eq.~\eqref{eq:app:Gamma_higher}]. In panel (a) the term $\tilde{\gamma}_2 \bm{S}_{21} \gamma_1 \tilde{\bm{S}}_{12}$ is shown. This process does not contribute to the current as the particle reteroreflected from the normal Andreev reflection on the superconducting side has the opposite spin to that needed for the following triplet Andreev reflection on the ferromagnetic side. As a result, this term needs an additional spin-flip process [marked by the green circle in Fig.~\ref{fig:app:normalisation_diagrams}(a)] to contribute to the current. Panel (b) shows the contribution $\tilde{\gamma}_2 \bm{S}_{22} \gamma_2 \tilde{\bm{S}}_{22}$ which is a first-order Andreev process in which a triplet pair gets totally reflected at the barrier. Consequently, this term does not contribute to the charge current either. Therefore, any combination of those two processes $(\tilde{\gamma}_i \gamma_{ii}^\prime)^l$ for $l \in \mathbb{N}$, cannot contribute to the charge current, making $n_2$ and $n_3$ not contributing either.

\begin{figure}[t!]
    \centering
    \includegraphics[width=\linewidth]{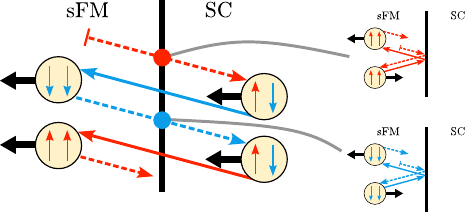}
    \caption{The diagrammatic representation of the process given by $\tilde{\gamma}_2 \gamma_{23}^\prime \tilde{\gamma}_3 \gamma_{32}^\prime$. The role of colors and line styles is the same as in Fig.~\ref{fig:lower_order_contributions}. The red and blue dot at the sFM/SC barrier denote the possibilities of multiple Andreev reflections preceeding the transmission of the hole from the sFM into the SC. A sketch of a lowest order reflection process is shown on the right hand side and connected to the dot by a grey line. Including these additional Andreev reflections yield the diagrammatic representation of the remaining terms in Eq.~\eqref{eq:app:n_ij_full}.}
    \label{fig:app:normalisation_second_order}
\end{figure}
\begin{table}[b!]
\centering
\caption{A list of equations and corresponding diagrams shown in figures throughout. Please note that the contributions listed in this table are the contributions to $\Gamma_\rS$.  The contributions to the current and the diagrams are obtained by multiplying the listed contributions by $\tilde{\gamma}_\rS$ from the right.}
\begin{tabular}{c|c}
Contribution & Diagram \\ \hline \hline
$r_\rS \gamma_\rS \tilde{r}_\rS$ & Fig.~\ref{fig:lower_order_contributions}(a) \\
$t_{\rS\uparrow} \gamma_\upup \tilde{t}_{\uparrow\rS}$ & Fig.~\ref{fig:lower_order_contributions}(b) \\
$\red{r_\rS\gamma_\rS\tilde{t}_{\rS\uparrow}} \blue{\tilde{\gamma}_\upup} \lila{r_\uparrow \gamma_\upup \tilde{t}_{\uparrow\rS}}$ & Fig.~\ref{fig:lower_order_contributions}(c) \\
$\green{t_{\rS\uparrow}\gamma_\upup \tilde{r}_\uparrow} \blue{\tilde{\gamma}_\upup} \orange{t_{\uparrow\rS}\gamma_\rS \tilde{r}_\rS}$ & Fig.~\ref{fig:lower_order_contributions}(d) \\
$\red{r_\rS \gamma_\rS \tilde{t}_{\rS\uparrow}} \blue{\tilde{\gamma}_\upup} \orange{t_{\uparrow\rS} \gamma_\rS \tilde{r}_\rS}$ & Fig.~\ref{fig:lower_order_contributions}(e) \\
$\green{t_{\rS\uparrow}\gamma_\upup\tilde{r}_\uparrow} \blue{\tilde{\gamma}_\upup} \lila{r_\uparrow \gamma_\upup \tilde{t}_{\uparrow\rS}}$ & Fig.~\ref{fig:lower_order_contributions}(f) \\
$\red{r_\rS \gamma_\rS \tilde{t}_{\rS\downarrow}} \blue{\tilde{\gamma}_\dodo t_{\downarrow\rS}\gamma_\rS \tilde{t}_{\rS\uparrow} \tilde{\gamma}_\upup} \orange{t_{\uparrow\rS} \gamma_\rS \tilde{r}_\rS}$ & Fig.~\ref{fig:fourth_order_Andreev_processes}(a) \\
$\green{t_{\rS\downarrow}\gamma_\dodo \tilde{r}_\downarrow} \blue{\tilde{\gamma}_\dodo t_{\downarrow\rS} \gamma_\rS \tilde{t}_{\rS\uparrow}\tilde{\gamma}_\upup} \orange{t_{\uparrow\rS}\gamma_\rS\tilde{r}_\rS}$ & Fig.~\ref{fig:fourth_order_Andreev_processes}(b) \\
$ \red{r_\rS \gamma_\rS \tilde{t}_{\rS\downarrow}} \blue{\tilde{\gamma}_\dodo t_{\downarrow\rS}\gamma_\rS\tilde{t}_{\rS\uparrow}\tilde{\gamma}_\upup} \lila{r_\uparrow \gamma_\upup \tilde{t}_{\uparrow\rS}}$ & Fig.~\ref{fig:fourth_order_Andreev_processes}(c) \\
$\green{t_{\rS\downarrow}\gamma_\dodo \tilde{r}_\downarrow} \blue{\tilde{\gamma}_\dodo t_{\downarrow\rS}\gamma_\rS \tilde{t}_{\rS\uparrow}\tilde{\gamma}_\upup} \lila{r_\uparrow \gamma_\upup \tilde{t}_{\uparrow\rS}}$ & Fig.~\ref{fig:fourth_order_Andreev_processes}(d) \\
$\red{r_\rS \gamma_\rS \tilde{t}_{\rS\downarrow}} \blue{\tilde{\gamma}_\dodo t_{\downarrow\rS} \gamma_\rS \tilde{t}_{\rS\uparrow} \tilde{\gamma}_\upup t_{\uparrow\rS} \gamma_\rS \tilde{t}_{\rS\downarrow} \tilde{\gamma}_\dodo} \orange{t_{\downarrow\rS} \gamma_\rS \tilde{r}_\rS}$ & Fig.~\ref{fig:sixth_order_Andreev_processes}(a) \\
$\green{t_{\rS\downarrow} \gamma_\dodo \tilde{r}_\downarrow} \blue{\tilde{\gamma}_\dodo t_{\downarrow\rS} \gamma_\rS \tilde{t}_{\rS\uparrow} \tilde{\gamma}_\upup t_{\uparrow\rS} \gamma_\rS \tilde{t}_{\rS\downarrow} \tilde{\gamma}_\dodo} \orange{t_{\downarrow\rS} \gamma_\rS \tilde{r}_\rS}$ & Fig.~\ref{fig:sixth_order_Andreev_processes}(b) \\
$\red{r_\rS \gamma_\rS \tilde{t}_{\rS\downarrow}} \blue{\tilde{\gamma}_\dodo t_{\downarrow\rS} \gamma_\rS \tilde{t}_{\rS\uparrow} \tilde{\gamma}_\upup t_{\uparrow\rS} \gamma_\rS \tilde{t}_{\rS\downarrow} \tilde{\gamma}_\dodo} \lila{r_\downarrow \gamma_\dodo \tilde{t}_{\downarrow\rS}}$ & Fig.~\ref{fig:sixth_order_Andreev_processes}(c) \\
$\green{t_{\rS\downarrow} \gamma_\dodo \tilde{r}_\downarrow} \blue{\tilde{\gamma}_\dodo t_{\downarrow\rS} \gamma_\rS \tilde{t}_{\rS\uparrow} \tilde{\gamma}_\upup t_{\uparrow\rS} \gamma_\rS \tilde{t}_{\rS\downarrow} \tilde{\gamma}_\dodo} \lila{r_\downarrow \gamma_\dodo \tilde{t}_{\downarrow\rS}}$ & Fig.~\ref{fig:sixth_order_Andreev_processes}(d)
\end{tabular}
\label{tab:relation_eq_diagram}
\end{table}

Next, we consider the contributions and diagrams originating from $n_{23}$. From Eq.~\eqref{eq:app:n_ij_full} it is evident that all contributions have similar structure and they differ only in the occurence of $n_2$ and/or $n_3$. Since we already discussed the contributions of the $n_i$'s we consider the lowest order contribution without $n_2$ or $n_3$ which is given as 
\begin{equation}
    \tilde{\gamma}_2 \gamma_{23}^\prime \tilde{\gamma}_3 \gamma_{32}^\prime \equiv \tilde{\gamma}_\upup t_{\uparrow\rS} \gamma_\rS \tilde{t}_{\rS \downarrow} \tilde{\gamma}_\dodo t_{\downarrow\rS} \gamma_\rS \tilde{t}_{\rS\uparrow}.
\end{equation}

\begin{figure}[t!]
    \centering
    \includegraphics[width=\linewidth]{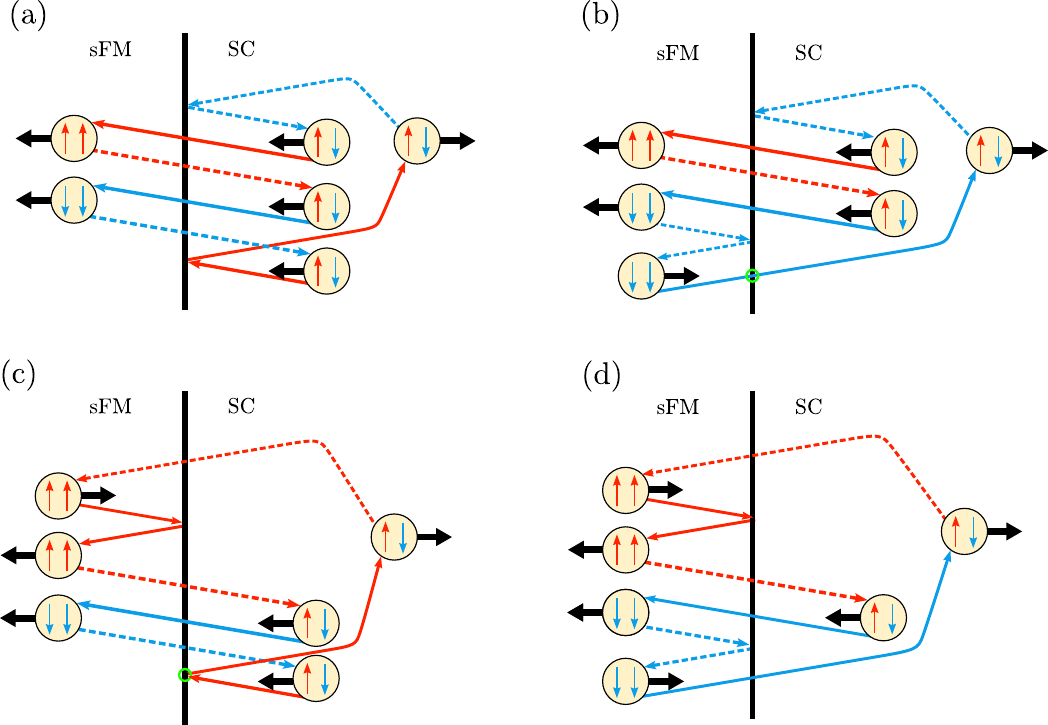}
    \caption{Andreev processes of order $t^4$ following from Eq.~\eqref{eq:contribution_fifth}. Table~\ref{tab:relation_eq_diagram} shows the relation between the equations and the diagrams. The notation is the same as for Fig.~\ref{fig:lower_order_contributions}. The contributions shown in panels (b)-(d) do not contribute to the supercurrent whereas panel (a) shows the lowest order contribution to the supercurrent leading to a second harmonic in the Josephson current.}
    \label{fig:fourth_order_Andreev_processes}
\end{figure}

To gain more insight, we consider a diagrammatic representation of this contribution shown in Fig.~\ref{fig:app:normalisation_second_order}. It is a fourth-order Andreev process yielding a second harmonic in the CPR. Considering the cases in which $n_2$ and/or $n_3$ must be taken into account [see the latter terms of Eq.~\eqref{eq:app:n_ij_full}], the corresponding diagrams for $n_{2/3}$ must be inserted. From the previous discussion it is clear that the contribution shown in Fig.~\ref{fig:app:normalisation_diagrams}(a) does not need to be taken into account whereas the contribution from Fig.~\ref{fig:app:normalisation_diagrams}(b) must be taken into account. For the ease of diagrammatic notation we added in Fig.~\ref{fig:app:normalisation_second_order} a red and a blue dot where the transmission of a hole from the sFM to SC occurs. These dots represent the processes indicated by the connecting lines. Thus, additional Andreev reflections take place before the hole is transmitted into the SC. Therefore, these contributions also yield a second harmonic contribution and only the magnitude is affected by these additional reflections.

Considering higher order contributions in Eq.~\eqref{eq:app:n_ij_full}, i.e., $k > 1$, yields higher harmonic contributions of order $2k$. This is because a multiplication of the different contributions inside the brackets of the sum in Eq.~\eqref{eq:app:n_ij_full} leads to a concatenation of the corresponding diagrams resulting in higher order contributions to the current.

\section{Lowest order contributions} \label{sec:app:diagrams}
\begin{figure}[t!]
	\centering
	\includegraphics[width=\linewidth]{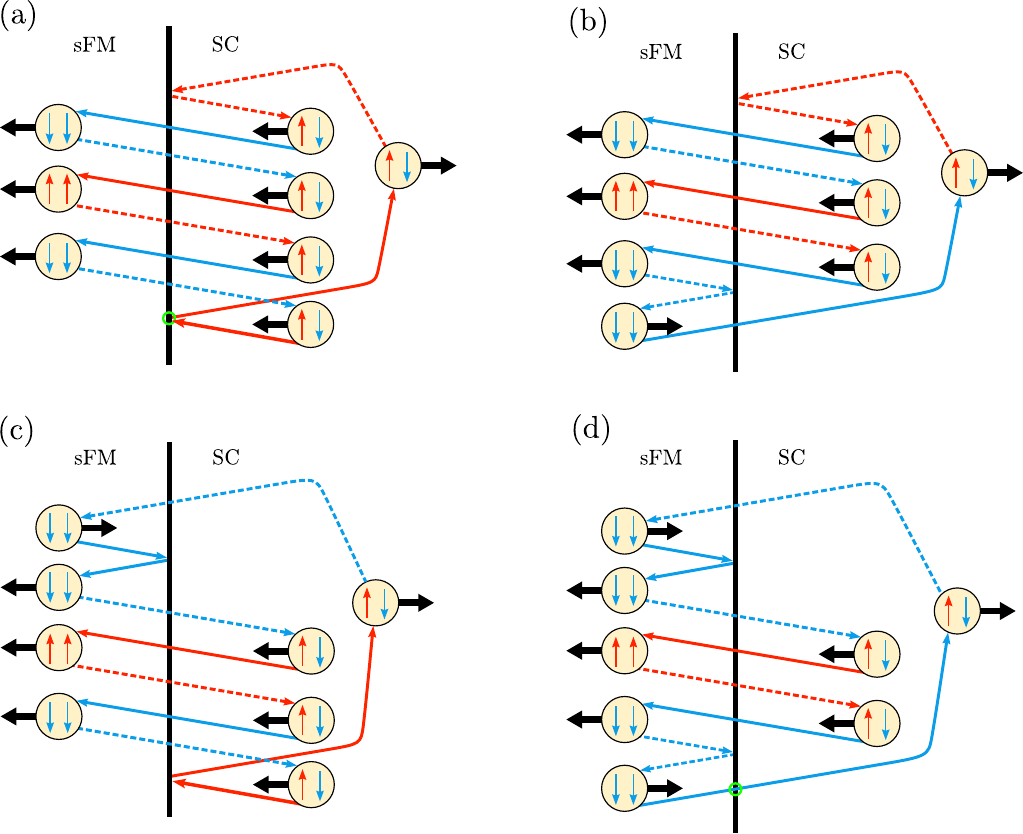}
	\caption{Andreev processes of order $t^6$ following from Eq.~\eqref{eq:contribution_seventh}. Table~\ref{tab:relation_eq_diagram} shows the relation between the equations and the diagrams. The notation is the same as for Fig.~\ref{fig:lower_order_contributions}. The contributions from panels (a) and (d) do not contribute to the supercurrent as a spin-flip process is required for their contribution. The contributions shown in (b) and (c) contribute to the supercurrent and lead to a second harmonic contribution in the CPR.}
	\label{fig:sixth_order_Andreev_processes}
\end{figure}

In the previous section, we showed that it is sufficient to consider the diagrams following from $\Gamma_1 - \Gamma_1^\mathrm{higher}$ [see Eqs.~\eqref{eq:app:Gamma_1} and \eqref{eq:app:Gamma_higher}]. In this appendix, we explicitly calculate the lowest order contributions by inserting the elementary scattering processes from Eq.~\eqref{eq:elementary_scatter_process} into $\Gamma_1 - \Gamma_1^\mathrm{higher}$, for which we use the S-matrix shown in Appendix~\ref{app:scatmat}. For simplicity, we omit the index $\ldots^\mathrm{R}$ but all reflection or transmission coefficients shown in the following are entries of the scattering matrix for the right interface. 
Consequently, the contributions to $\Gamma_1 - \Gamma_1^\mathrm{higher}$ read
\begingroup
\allowdisplaybreaks
\begin{align}
    \gamma_{11}^\prime &= r_\rS \gamma_\rS \tilde{r}_\rS + t_{\rS\uparrow} \gamma_\upup \tilde{t}_{\uparrow\rS} + t_{\rS\downarrow} \gamma_\dodo \tilde{t}_{\downarrow \rS}, \label{eq:contribution_first} \\
    \begin{split}
        \gamma_{12}^\prime \tilde{\gamma}_2 \gamma_{21}^\prime &= r_\rS\gamma_\rS\tilde{t}_{\rS\uparrow}\tilde{\gamma}_\upup r_\uparrow \gamma_\upup \tilde{t}_{\uparrow \rS} \\
        &\quad + t_{\rS\uparrow}\gamma_\upup \tilde{r}_\uparrow \tilde{\gamma}_\upup t_{\uparrow \rS}\gamma_\rS \tilde{r}_\rS\\
        &\quad + r_\rS \gamma_\rS \tilde{t}_{\rS\uparrow} \tilde{\gamma}_\upup t_{\uparrow \rS} \gamma_\rS \tilde{r}_\rS  \\
        &\quad + t_{\rS\uparrow}\gamma_\upup\tilde{r}_\uparrow\tilde{\gamma}_\upup r_\uparrow \gamma_\upup \tilde{t}_{\uparrow \rS} \label{eq:contribution_third},
    \end{split} 
\end{align}
\endgroup
\begin{align}\label{eq:contribution_fifth}
        \gamma_{13}^\prime \tilde{\gamma}_3 \gamma_{32}^\prime \tilde{\gamma}_2 \gamma_{21}^\prime &= r_\rS \gamma_\rS \tilde{t}_{\rS\downarrow} \tilde{\gamma}_\dodo t_{\downarrow \rS}\gamma_\rS \tilde{t}_{\rS\uparrow} \tilde{\gamma}_\upup t_{\uparrow \rS} \gamma_\rS \tilde{r}_\rS\\
        &\quad + t_{\rS\downarrow}\gamma_\dodo \tilde{r}_\downarrow \tilde{\gamma}_\dodo t_{\downarrow \rS} \gamma_\rS \tilde{t}_{\rS\uparrow}\tilde{\gamma}_\upup t_{\uparrow \rS}\gamma_\rS\tilde{r}_\rS\nonumber \\
        &\quad + r_\rS \gamma_\rS \tilde{t}_{\rS\downarrow}\tilde{\gamma}_\dodo t_{\downarrow \rS}\gamma_\rS\tilde{t}_{\rS\uparrow}\tilde{\gamma}_\upup r_\uparrow \gamma_\upup \tilde{t}_{\uparrow \rS}\nonumber \\
        &\quad + t_{\rS\downarrow}\gamma_\dodo \tilde{r}_\downarrow \tilde{\gamma}_\dodo t_{\downarrow \rS}\gamma_\rS \tilde{t}_{\rS\uparrow}\tilde{\gamma}_\upup r_\uparrow \gamma_\upup \tilde{t}_{\uparrow \rS},\nonumber 
\end{align}
\begin{align}
        \gamma_{13}^\prime &\tilde{\gamma}_3 \gamma_{32}^\prime\tilde{\gamma}_2 \gamma_{23}^\prime \tilde{\gamma}_3 \gamma_{31}^\prime= \nonumber\\
        \label{eq:contribution_seventh}
        &=r_\rS \gamma_\rS \tilde{t}_{\rS\downarrow} \tilde{\gamma}_\dodo t_{\downarrow\rS} \gamma_\rS \tilde{t}_{\rS\uparrow} \tilde{\gamma}_\upup t_{\uparrow\rS} \gamma_\rS \tilde{t}_{\rS\downarrow} \tilde{\gamma}_\dodo t_{\downarrow\rS} \gamma_\rS \tilde{r}_\rS \\
        &+ t_{\rS\downarrow} \gamma_\dodo \tilde{r}_\downarrow \tilde{\gamma}_\dodo t_{\downarrow\rS} \gamma_\rS \tilde{t}_{\rS\uparrow} \tilde{\gamma}_\upup t_{\uparrow\rS} \gamma_\rS \tilde{t}_{\rS\downarrow} \tilde{\gamma}_\dodo t_{\downarrow\rS} \gamma_\rS \tilde{r}_\rS\nonumber \\
        &+ r_\rS \gamma_\rS \tilde{t}_{\rS\downarrow} \tilde{\gamma}_\dodo t_{\downarrow\rS} \gamma_\rS \tilde{t}_{\rS\uparrow} \tilde{\gamma}_\upup t_{\uparrow\rS} \gamma_\rS \tilde{t}_{\rS\downarrow} \tilde{\gamma}_\dodo r_\downarrow \gamma_\dodo \tilde{t}_{\downarrow\rS}\nonumber \\ 
        &+ t_{\rS\downarrow} \gamma_\dodo \tilde{r}_\downarrow \tilde{\gamma}_\dodo t_{\downarrow\rS} \gamma_\rS \tilde{t}_{\rS\uparrow} \tilde{\gamma}_\upup t_{\uparrow\rS} \gamma_\rS \tilde{t}_{\rS\downarrow} \tilde{\gamma}_\dodo r_\downarrow \gamma_\dodo \tilde{t}_{\downarrow\rS}.\nonumber
\end{align}
The remaining contributions of second, fourth, and sixth order in transmission follow immediately from Eqs.~\eqref{eq:contribution_third}, \eqref{eq:contribution_fifth}, and \eqref{eq:contribution_seventh} by exchanging $\uparrow$ and $\downarrow$ in the corresponding equation. The diagrams following from Eqs.~\eqref{eq:contribution_first} and \eqref{eq:contribution_third} are shown in Fig.~\ref{fig:lower_order_contributions}. The diagrams of fourth and sixth order in transmission are shown in Figs.~\ref{fig:fourth_order_Andreev_processes} and \ref{fig:sixth_order_Andreev_processes}. Higher order processes are included in $\Gamma_1^\mathrm{higher}$.

\bibliography{references}
\end{document}